\documentclass[aps,reprint,twocolumn,floatfix,preprintnumbers,superscriptaddress,amsmath,amssymb]{revtex4-2}
\usepackage{xcolor,bm,lineno,multirow,float,times}
\usepackage{physics}
\usepackage{rotating}
\usepackage{verbatim}
\usepackage{graphicx,tabularx}
\usepackage[multiple]{footmisc}
\usepackage[sort&compress]{natbib}
\usepackage[T1]{fontenc}
\usepackage{soul}

\usepackage[%
  colorlinks=true,
  urlcolor=blue,
  linkcolor=blue,
  citecolor=blue
]{hyperref}

\newcommand{\blue}[1]{{\color{black} #1}}

\def\eg{{e.g. }}
    
\begin{document}

\title{Anticollinear order and degeneracy lifting in square lattice antiferromagnet LaSrCrO$_4$}

\author{Jing Zhou}
\affiliation{Institute of Physics, Chinese Academy of Sciences, Beijing 100190, China}
\affiliation{University of the Chinese Academy of Sciences, Beijing 100049, China}

\author{Guy Quirion}
\affiliation{Department of Physics and Physical Oceanography, Memorial University of Newfoundland, St. John’s, Canada A1B 3X7}

\author{Jeffrey A. Quilliam}
\affiliation{Institute Quantique, Départment de Physique, and RQMP, Université de Sherbrooke, Sherbrooke, Québec, Canada J1K 2R1 }

\author{Huibo Cao}
\affiliation{Neutron Scattering Division, Oak Ridge National Laboratory, Oak Ridge, TN 37831, USA}

\author{Feng Ye}
\affiliation{Neutron Scattering Division, Oak Ridge National Laboratory, Oak Ridge, TN 37831, USA}

\author{Matthew B. Stone}
\affiliation{Neutron Scattering Division, Oak Ridge National Laboratory, Oak Ridge, TN 37831, USA}

\author{Qing Huang}
\affiliation{Department of Physics and Astronomy, University of Tennessee, Knoxville, TN 37996, USA}

\author{Haidong Zhou}
\affiliation{Department of Physics and Astronomy, University of Tennessee, Knoxville, TN 37996, USA}

\author{Jinguang Cheng}
\affiliation{Institute of Physics, Chinese Academy of Sciences, Beijing 100190, China}

\author{Xiaojian Bai}
\altaffiliation{Present Adress: Neutron Scattering Division, Oak Ridge National Laboratory, Oak Ridge, TN 37831, USA}
\affiliation{School of Physics, Georgia Institute of Technology, Atlanta, GA 30332, USA}

\author{Martin Mourigal}
\affiliation{School of Physics, Georgia Institute of Technology, Atlanta, GA 30332, USA}

\author{Yuan Wan}
\email[]{yuan.wan@iphy.ac.cn}
\affiliation{Institute of Physics, Chinese Academy of Sciences, Beijing 100190, China}
\affiliation{University of the Chinese Academy of Sciences, Beijing 100049, China}
\affiliation{Songshan Lake Materials Laboratory, Dongguan, Guangdong 523808, China}

\author{Zhiling Dun}
\email[]{dun@iphy.ac.cn}
\affiliation{Institute of Physics, Chinese Academy of Sciences, Beijing 100190, China}
\affiliation{School of Physics, Georgia Institute of Technology, Atlanta, GA 30332, USA}

\date{\today}

\begin{abstract}
We report the static and dynamic magnetic properties of LaSrCrO$_4$, a seemingly canonical spin-3/2 square-lattice antiferromagnet that exhibits frustration between magnetic layers -- owing to their AB stacking -- and offers a rare testbed to investigate accidental-degeneracy lifting in magnetism. Neutron diffraction experiments on single-crystal samples uncover a remarkable anticollinear magnetic order below $T_N$ = 170 K characterized by a N\'eel arrangement of the spins within each layer and an orthogonal arrangement between adjacent layers. To understand the origin of this unusual magnetic structure, we analyze the spin-wave excitation spectrum by means of inelastic neutron scattering and bulk measurements. A spectral gap of 0.5 meV, along with a spin-flop transition at 3.2\,T, reflect the energy scale associated with the degeneracy-lifting. A minimal model to explain these observations requires both a positive biquadratic interlayer exchange and dipolar interactions, both of which are on the order of 10$^{-4}$ meV, only a few parts per million of the dominant exchange interaction $J_1 \approx 11$ meV. These results provide direct evidence for the selection of a non-collinear magnetic structure by the combined effect of two distinct degeneracy lifting interactions.
\end{abstract} 

\maketitle

\textit{Introduction.} 
The emergence of accidental ground state degeneracy and its lifting are central to our understanding of frustrated magnetism~\cite{Ramirez1994,Moessner2001,Chalker2011}. The interplay between exchange interactions and lattice geometry often result in a family of accidentally degenerate ground states that are unrelated by symmetry. The degeneracy is then lifted either by subleading interactions, \eg magnetic dipolar interaction~\cite{Melko2001,Ruff2005,Moller2009,Chern2011}, magnetoelastic coupling~\cite{Tchernyshyov2002a,Tchernyshyov2002b}, etc.; or by fluctuations that normally work against ordering, \eg quenched disorder, thermal or quantum fluctuations, through the ``order by disorder (ObD)" mechanism ~\cite{Tessman1954,Villain1979,Shender1982,Kawamura1984,Henley1989,Savary2012,Smirnov2017}. The diverse degeneracy lifting mechanisms can stabilize a host of magnetic orders in materials with similar structures and chemical compositions \cite{Hallas2018, Dun2016}, and their competition offers flexible tunablity in and out-of-equilibrium~\cite{Wan2017,Wan2018}. Yet, experimentally revealing the degeneracy lifting mechanism is a challenging task due to the minuscule energy scales,  sometimes in the one part per million of the dominant exchange interaction, associated with these subleading interactions and/or the ObD effects~\cite{Ross2014a}.


\begin{figure}[tbp!]
	\linespread{1}
	\par
	\begin{center}
		\includegraphics[width= \columnwidth]{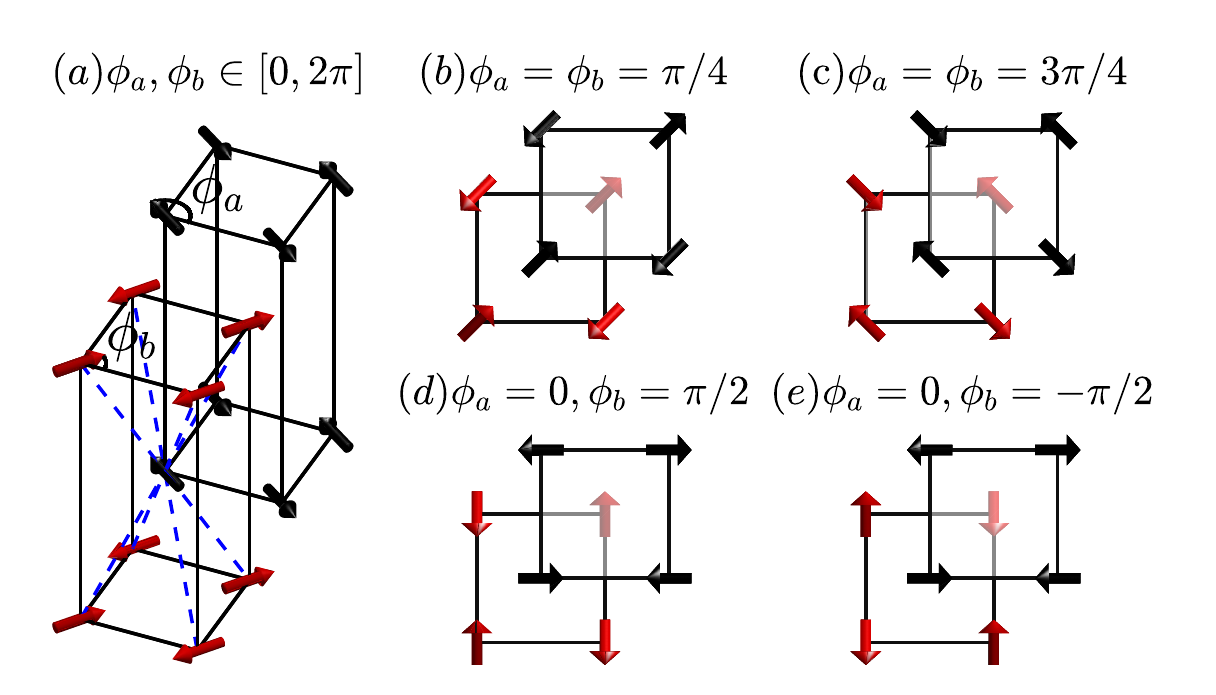}
	\end{center}
	\par
	\caption{\label{Fig:spins}(a) The quasi-2D square lattice antiferromagnet with AB stacking comprises of two sublattices (dubbed A and B), each hosting a 3D N\'{e}el order. The two N\'{e}el vectors are decoupled at the mean field level owing to the frustrated interlayer coupling. An easy-plane single-ion anisotropy forces the N\'{e}el vectors to be in the crystallographic $ab$ plane, which are then parametrized by their respective azimuthal angles ($\phi_a$, $\phi_b$). (b) Collinear spin structure with $\phi_a = \phi_b = \frac{\pi}{4}$, observed in La$_2$CuO$_4$~\cite{Vaknin1987},  Sr$_2$CuO$_2$Cl$_2$~\cite{You2000}, LaSrFeO$_4$~\cite{Qureshi2013}, and La$_2$CoO$_4$ in the orthorhombic phase~\cite{Yamada1989}. (c) Collinear spin structure with $\phi_a = \phi_b = \frac{3\pi}{4}$ for La$_2$NiO$_4$~\cite{Aeppli1988} and possibly La$_2$CoO$_4$ in the low temperature tetragonal phase~\cite{Yamada1989,Gardner1997}.  (d) Anticollinear state with $\phi_a = 0$,  $\phi_b = \frac{\pi}{2}$, for LaSrCrO$_4$ reported in this work. (e) The other symmetry-inequivalent anticollinear order with $\phi_a=0$, $\phi_b=-\frac{\pi}{2}$.}

\end{figure}

The quasi two-dimensional (2D) square lattice Heisenberg antiferromagnet with AB stacking is a prominent model system to illustrate the diverse degeneracy lifting mechanisms and the wealth of resulting magnetic orders~\cite{Henley1989,Yildirim1994a,Yildirim1994b,Yildirim1996}. The antiferromagnetic intra-layer exchange interaction stablize a 2D N\'eel order in each layer. However, the inter-layer exchange interactions are frustrated owing to the AB stacking. Consequently, the N\'{e}el vectors in two adjacent layers remain decoupled at the mean field level, thereby giving rise to a continuous manifold of accidentally degenerate ground states, which can then be selected by various mechanisms. In particular, the thermal and quantum fluctuations stabilize the collinear arrangement of N\'{e}el vectors through the ObD mechanism, whereas the quenched disorder favors anticollinear orders where the N\'eel vectors are orthogonal~\cite{Henley1989}.

\begin{figure}
	\linespread{1}
	\par
	\begin{center}
		\includegraphics[width= \columnwidth ]{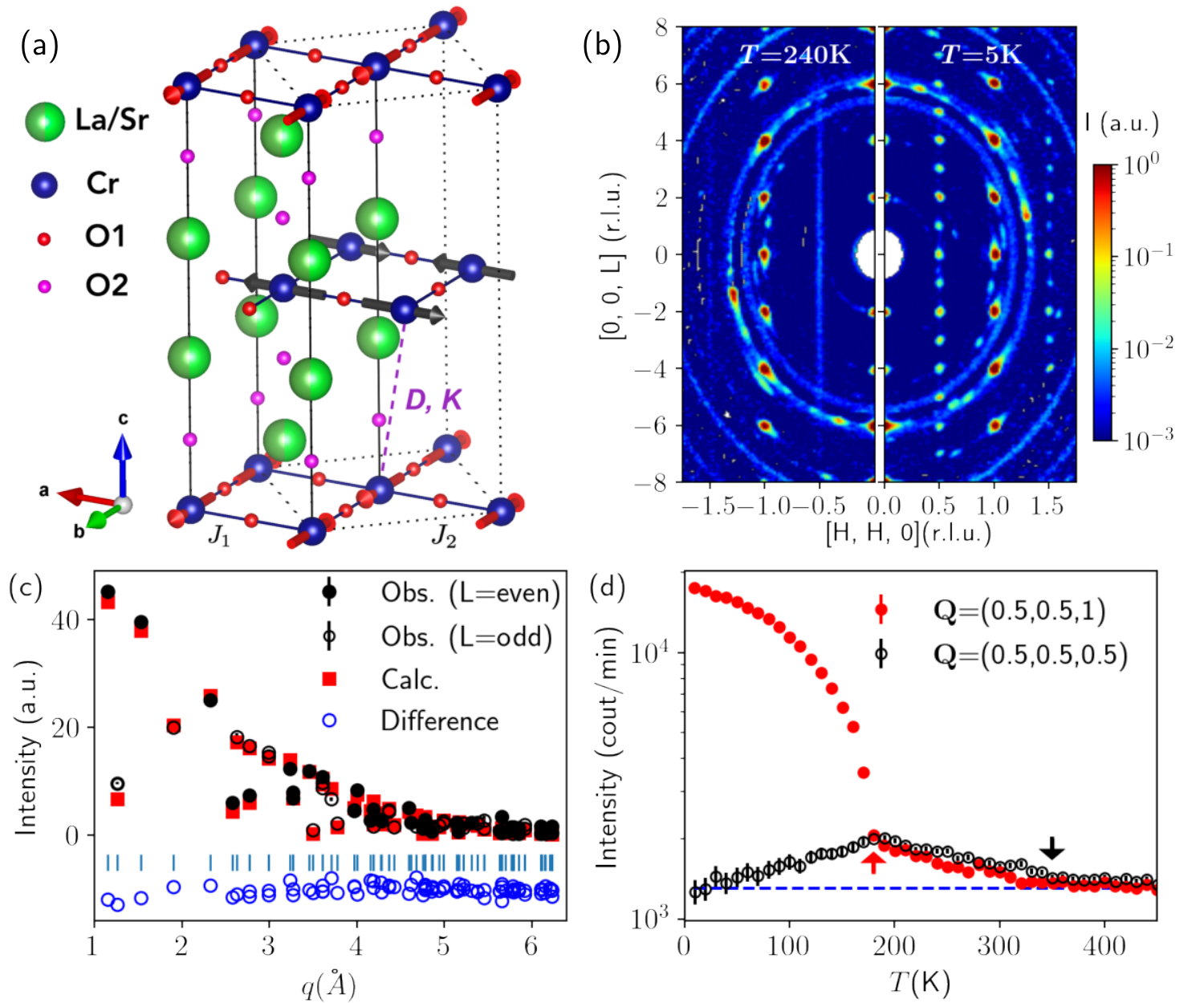}
	\end{center}
	\par
	\caption{\label{Fig:Magstru} (a) Nuclear and magnetic unit cells (represented by solid and dotted lines) of LSCrO. Colored spheres present different atoms and red/black arrows presents Cr$^{3+}$ spins that are orthogonal between adjacent layers. Spin interactions $J_1, J_2, D, K$ in Eqs.~\eqref{eq:H} and \eqref{eq:H_interlayer} are labeled for selective Cr-Cr bonds. (b) Elastic neutron scattering patterns in the (HHL) plane, measured on SEQUOIA (Spallation Neutron Source, Oak Ridge National Laboratory, Ref.~\cite{Granroth2010}) at $T$ = 240 K, and 5 K, respectively. Intensities are integrated within $\pm$0.1 reciprocal-lattice unit (r.l.u.) in the [$\mathrm{K\bar{K}}0$] direction. (c) Rietveld refinement of the magnetic reflections collected on HB3a (High Flux Isotope Reactor, ORNL, Ref.~\cite{Chakoumakos2011}) at 4\,K based on the magnetic structure shown in (a). (d) Temperature dependence of the magnetic diffuse scattering intensity at $\mathbf{Q} =  (0.5, 0.5, 0.5)$ and magnetic Bragg peak intensity at $\mathbf{Q} = (0.5, 0.5, 1)$. The onset temperatures and 2D and 3D magnetic ordering  are indicated by the arrows. }
\end{figure}

Experimentally, such interlayer frustration exists in a large family of transition metal oxides with a layered perovskite structure of the K$_2$NiF$_4$ type [space group $I4/mmm$, Fig.~\ref{Fig:Magstru}(a)] and easy-plane single-ion anisotropy. Focusing on simple systems without secondary magnetic lattices or electron/hole doping, including La$_2$MO$_4$ (M = Cu~\cite{Vaknin1987},  Ni~\cite{Aeppli1988,  Wang1992},  Co~\cite{Yamada1989, Gardner1997, Babkevich2010}),  LaSrFeO$_4$~\cite{Qureshi2013} and Sr$_2$CuO$_2$Cl$_2$~\cite{You2000}, all of these compounds exhibit collinear orders without exception [Fig.~\ref{Fig:spins}(b)(c)]. In La$_2$MO$_4$ (M = Cu, Ni, Co), the orthorhombic lattice distortion lifts the degeneracy and stabilizes the collinear order~\cite{Aeppli1988, Yamada1989, AvinashSingh1990}. In LaSrFeO$_4$ and Sr$_2$CuO$_2$Cl$_2$, the lattice distortion is absent; the degeneracy lifting mechanism is less clear though thermal or quantum fluctuations are likely responsible~\cite{Henley1989,Yildirim1996}.

In this work, we investigate a much less characterized member of this material family, LaSrCrO$_4$ (LSCrO)~\cite{ Aso1978,Madrid1994, Kao2015}. Using neutron scattering measurements on a single crystal sample, we reveal a striking \emph{anticollinear} magnetic ground state [Fig.~\ref{Fig:spins}(d)] that is distinct from all the compounds mentioned above. Combining theoretical analysis with various experimental measurements, we show that the magnetic dipolar interaction and the biquadratic spin exchange interaction, both on the order of 10$^{-5}$ (10 ppm) of the main exchange interaction $J_1$, are responsible for lifting the degeneracy and stabilizing the anticollinear state in this material. Our results thus establish LSCrO as a rare example where the degeneracy lifting interactions with minuscule energy scales can be exposed unambiguously.
   
\textit{Anticollinear order.}  We grow for the first time centimeter-sized single crystals of LSCrO via the floating zone technique \cite{Supplemental}. X-ray and neutron diffraction measurements confirm that it crystallizes in the tetragonal space group $I4/mmm$ at room temperature with lattice constants $a=b=3.87218(3)$ {\AA}, $c=12.516(1)$ {\AA} [Fig.~\ref{Fig:Magstru}(a)],  consistent with previous reports \cite{Aso1978, Kao2015}.  By using Rietveld refinement of the nuclear Bragg peaks measured at various temperatures, we found no structural phase transitions down to 4\,K. Similar to other quasi-2D system \cite{PSchumann2008}, the magnetic ordering in LSCrO occurs in two steps. At temperatures below 350 K, short-ranged 2D N\'eel order develops gradually, evidenced by the increasing magnetic scattering intensities at the $M$-point of the square lattice Brillouin zone, which are diffuse along the L direction [Fig.~\ref{Fig:Magstru}(b) and (d)].  

Below $T_N$ = 170\,K, the diffuse scattering quickly concentrates into sharp magnetic Bragg peaks at wave vectors $\mathbf{Q} = (H+\frac{1}{2}, K+\frac{1}{2}, L)$ in reciprocal space [Fig. \ref{Fig:Magstru}(b)], pinpointing a three-dimensional (3D) ordering of Cr$^{3+}$ spins. Interestingly, magnetic Bragg peaks are observed at $\mathbf{Q}$ with \emph{both} even and odd $L$ [Fig.~\ref{Fig:Magstru}(b)(c)]. This observation can not be explained by the 3D N\'eel order with a single ordering wave vector, where the magnetic structure factor would be extinct at either even or odd $L$~\cite{Li2005}. In other words, the spins in adjacent planes cannot be strictly collinear. A Rietveld refinement of magnetic Bragg peak intensities collected at 4\,K indicates that the magnetic structure is best fit by a 2-$k$ model [$k_1$ = (1/2, 1/2, 0) and $k_2$ = (1/2, -1/2, 0)] with the ordered moment of 2.25(2)$\mu_B$/Cr$^{3+}$, characterized by the magnetic space group P$_C$4$_2$/ncm~\cite{Perez-Mato2015a}. The resulting magnetic structure is shown in Fig.~\ref{Fig:Magstru}(a), which is identical to the anticollinear structure shown in Fig.~\ref{Fig:spins}(d). 

\begin{figure}
	\linespread{1}
	\par
	\begin{center}
		\includegraphics[width= \columnwidth ]{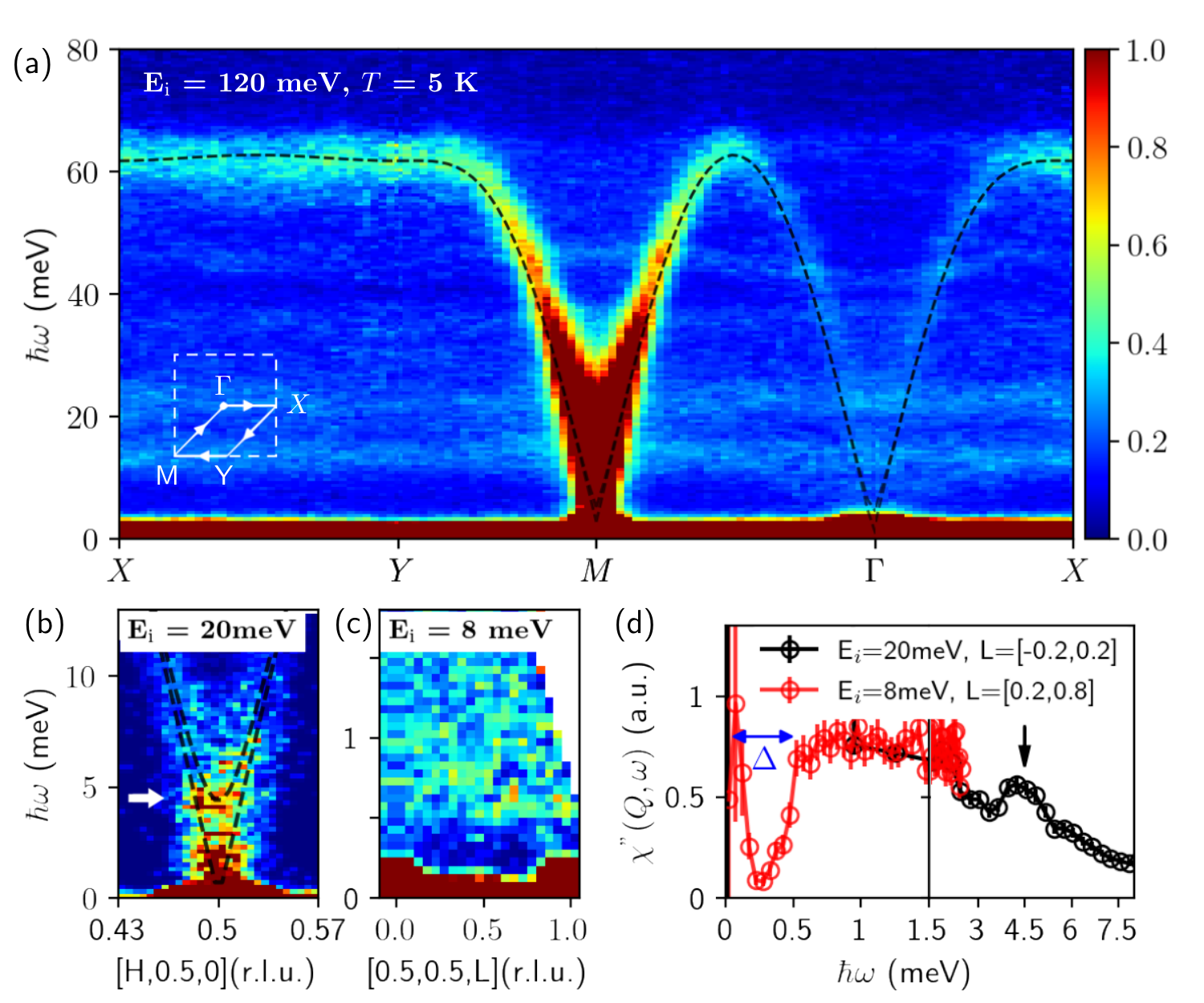}
	\end{center}
	\par
	\caption{\label{Fig:INS} (a) Spin wave excitations along high symmetry directions in the 2D Brillouin zone (inset) measured on SEQUOIA using $E_i$ = 120 meV.  Data are integrated within $H, K$= $\pm0.2$ and $L=\pm$8 r.l.u. The flat modes near 10 and 20 meV are optical phonons at high $L$ values. (b) Dispersion along $H$ near the $M$-point of the 2D Brillouin zone measured at $E_i$ = 20 meV. Data integration range is $H(K)$= $\pm0.03$ and $L=\pm$0.15 r.l.u. Dashed lines in (a) (b) represent best fit to Eq. \ref{eq:H} from LSWT. (c) Dispersion along $[0.5,0.5,L]$ measured at $E_i$ = 8 meV. Data integration range is $H,K$= $\pm0.02$ r.l.u. (d) Dynamic magnetic susceptibility at the $M$-point, obtained by integrating the data in (c). All data shown in this figure are collected at $T$ = 5\,K and symmetrized according to the $D_{4h}$ point group symmetry of the Cr$^{3+}$ site. }
\end{figure}

\textit{Spin wave excitations.}
We investigate the low temperature magnetic excitation spectrum of LSCrO using  time-of-flight neutron spectroscopy with various neutron incident energies ($E_{i}$) \cite{Supplemental}. Fig.~\ref{Fig:INS}(a) shows the overall energy-momentum dependence of the measured dynamic structure factor $S(q, \omega)$ along the high symmetry directions of the 2D Brillouin zone, where the scattering intensities are integrated along the $L$ direction. An intense and dispersive spin wave band emanates from the $M$-point. Its intensities gradually diminish when moving to the $\Gamma$ point. The spin wave shows almost no dispersion from the $X$ point to the $Y$ point, suggesting that further-neighbor exchange couplings~\cite{Headings_2010} and quantum anomaly effects ~\cite{DallaPiazza2015} are small. 

Using a lower incident energy, $E_i = 20$ meV, and therefore better energy resolution, we identify an energy gap of 4.5(1)\,meV in the $M$-point spectrum [indicated by arrows in Fig.~\ref{Fig:INS} (b)(d)]. We attribute this gap to the weak, easy-plane single-ion anisotropy of the Cr$^{3+}$ moments.

Given the large spin carried by the Cr$^{3+}$ ions [electron configuration $t_{2g}^3$, $S=3/2$], we expect that the observed spectrum can be understood in terms of the linear spin wave theory (LSWT). We find that the following minimal model Hamiltonian, which includes the first ($J_1$) and the second neighbor ($J_2$) exchange interactions, as well as an easy-plane single-ion anisotropy ($A$), can well describe the in-plane dispersion of the spin wave within the LSWT framework: 
\begin{equation}\label{eq:H}
\mathcal{H}_0 =  J_{1}\sum_{\langle i j\rangle_{1}}  \mathbf{S}_{i} \cdot \mathbf{S}_{j} + J_{2} \sum_{\langle i j\rangle_{2}} \mathbf{S}_{i} \cdot \mathbf{S}_{j} + A \sum_i ({S}_{i}^{z})^2, 
\end{equation}
where the summation $\langle i j\rangle_{n}$ runs over $n$-th neighbor spin pairs. We attain the best fit [dashed black lines in Fig. \ref{Fig:INS}(a)(b)] with $J_1 = 10.6(1)$\,meV, $J_2 = 0.16(6)$\,meV, $A = 0.05(1)$\, meV. The energy scale of the $J_1$ exchange is comparable to the onset temperature for the short-ranged 2D N\'{e}el order.  

Finally, we examine the low energy dispersion along the $L$ direction at the $M$-point with the best energy resolution obtained at $E_i = 8$ meV [Fig. \ref{Fig:INS}(c)]. Remarkably, the spectrum is gaped throughout.  As the gaps do not show discernible $L$-dependence, we conclude that the interlayer couplings between Cr$^{3+}$ spins of adjacent layers are smaller than the instrument resolution [$>0.1$\,meV]. By integrating $L$ in Fig. \ref{Fig:INS}(c) and avoiding regions where there is inelastic leakage from magnetic Bragg peaks, we obtain the energy dependence of dynamic susceptibility $\chi''(\bf{Q},\omega)$, which clearly reveals a second, much smaller gap $\Delta \approx $ 0.5(1) meV [Fig. \ref{Fig:INS}(d)].

The weak interlayer coupling is expected given the relative low 3D ordering temperature, $k_BT_N/[J_1 S(S+1)] = 0.391$. As a crude estimate, we neglect the small easy-plane anisotropy and utilize the published ordering temperatures of the quasi-2D Heisenberg model as determined by quantum Monte Carlo simulations~\cite{Majlis1992,Yasuda2005}. We estimate the interlayer coupling is in the range of $10^{-6}$ meV to $10^{-3}$ meV \cite{Supplemental}.

\textit{Interlayer couplings.} 
While the minimal model Eq.~\eqref{eq:H} can produce the in-plane dispersion of the spin wave, it is silent on the origin of the 3D magnetic structure. We now discuss the interlayer couplings that can stabilize the anticollinear state of LSCrO.

To set the stage, we determine the symmetry-allowed couplings between the N\'{e}el vectors associated with the two sublattices. The single ion anisotropy forces the N\'{e}el vectors to lie in the plane. We parametrize the orientation of the N\'{e}el vector in the sublattice A/B by the azimuthal angle $\phi_a$/$\phi_b$, respectively [Fig.~\ref{Fig:spins}(a)]. The interaction energy can be expanded as Fourier series of $\phi_{a,b}$. Up to the 4th order harmonics, our symmetry analysis yields three algebraically independent coupling terms \cite{Supplemental}:  $-\sin(\phi_a+\phi_b)$, $-\cos(4\phi_a)-\cos(4\phi_b)$, and $\cos(2\phi_a-2\phi_b)$. The signs at the front are needed to energetically favor the anticollinear state, i.e. $\phi_a = 0$, $\phi_b = \pi/2$ (and symmetry-related configurations). Each term admits a physical interpretation: The first term arises from the magnetic (pseudo) dipolar interaction; the second describes an in-plane, four-fold symmetric single ion anisotropy; the last comes from the biquadratic exchange interaction. 

Stabilizing the anticollinear order found in LSCrO requires the combination of either (a) dipolar interaction and biquadratic exchange or (b) dipolar interaction and single-ion anisotropy. Note the combination of the biquadratic exchange and the single-ion anisotropy does not fully lift the accidental degeneracy --- it admits another, \emph{symmetry-inequivalent} anticollinear state $\phi_a = 0$, $\phi_b = -\pi/2$ [Fig.~\ref{Fig:spins}(e)] in addition to the state observed in LSCrO.

Among the two possible combinations, we find the first can produce the correct spin flop transition observed in LSCrO (see below). We thus arrive at the following minimal Hamiltonian for the interlayer coupling:
\begin{align}\label{eq:H_interlayer}
\mathcal{H}' &= \sum_{\langle ij\rangle_3} D (\mathbf{S}_{i} \cdot \mathbf{S}_{j}-3 (\mathbf{S}_{i}\cdot\hat{n}_{ij}) (\mathbf{S}_{j}\cdot\hat{n}_{ij})) \nonumber\\
& + K(\mathbf{S}_{i} \cdot \mathbf{S}_{j})^{2},
\end{align}
where the summation is over all third-neighbor pairs. $\hat{n}_{i j}$ is the unit vector pointing from site $i$ to site $j$. $D>0$ and $K>0$ are strength of the dipolar and biqudratic couplings, respectively.  

\begin{figure}
	\linespread{1}
	\par
	\begin{center}
		\includegraphics[width= \columnwidth ]{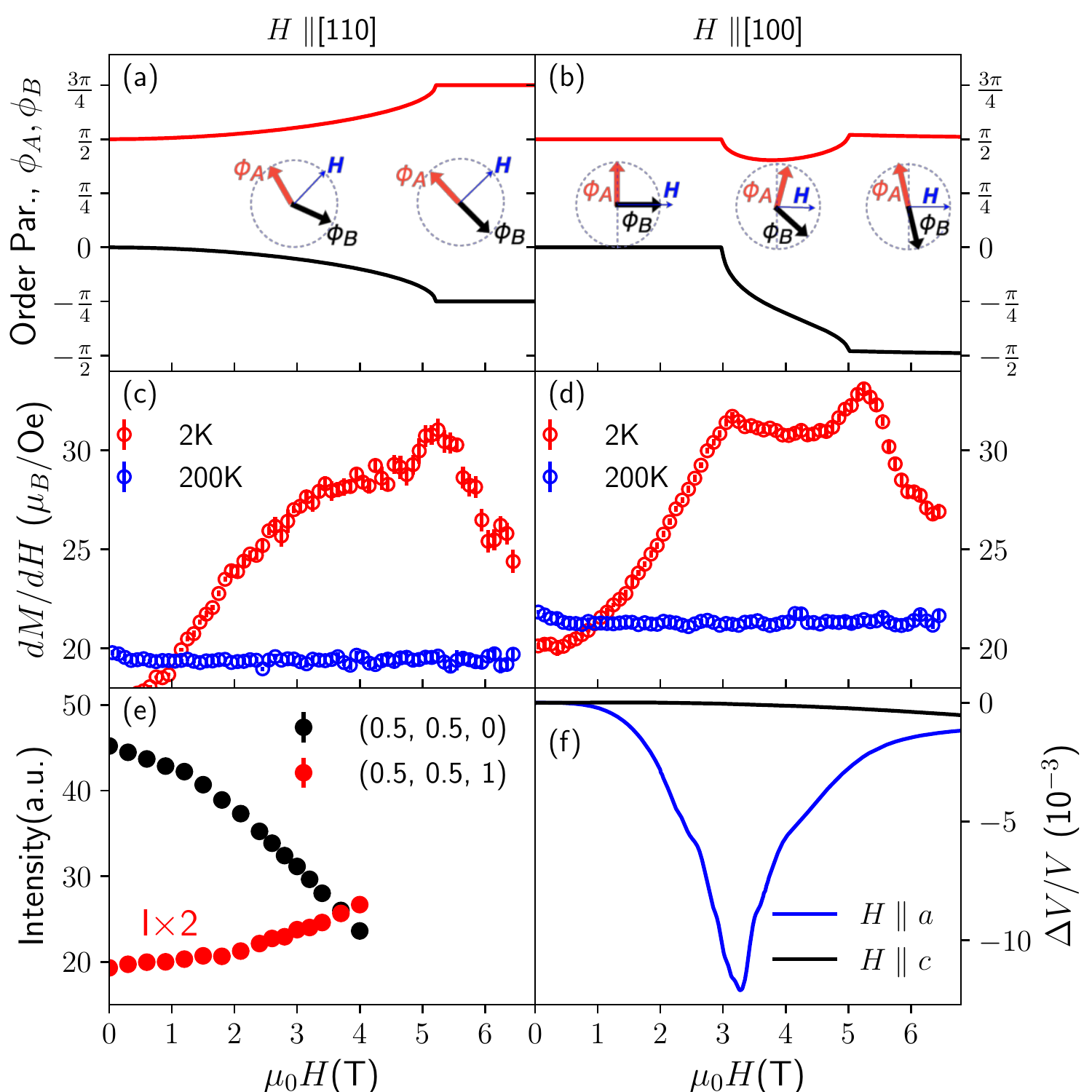}
	\end{center}
	\par
	\caption{\label{Fig:Spinflop} (a)(b) The azimuthal angle of the N\'{e}el vectors, $\phi_a$ and $\phi_b$, as a function of the applied magnetic field $H$ along (a) [110] and (b) [100] directions. Inset: configurations of the N\'{e}el vectors at selective fields. (c)(d) Differential magnetization at 2\,K and 200\,K, measured with $H$ applied along (c) [110] and (d) [100] directions.  (e) Field dependence of the magnetic Bragg peak intensities at $\bf{Q}$ = (0.5,0.5,0) and $\bf{Q}$ = (0.5,0.5,1), measured at $T$ = 2 K on the CORELLI diffuse scattering spectrometer (SNS, ORNL, Ref.~\cite{ye2018}) with magnetic field applied along the [1$\bar{1}$0] direction. (f) Field dependence of the relative velocity variation of the transverse mode propagating along x-axis and polarized along the y-axis $V_{LxPy}$, measured at $T$ = 2 K for $H$ along the $a$-axis (black curve) and along the $c$-axis (blue curve). }
\end{figure}

\textit{Spin flop transitions.} 
We now turn to the experimental test of the model Eq.~\eqref{eq:H_interlayer}. A sensitive diagnostic for the interlayer coupling is the spin flop transition driven by a magnetic field applied within $ab$ plane. The Zeeman coupling favors the N\'{e}el vectors to be perpendicular to the field in each layer. When the field is sufficiently strong, this effect can overcome the dipolar/biquadratic interactions and stabilize a collinear state. The resulting evolution from an anticollinear to a collinear magnetic structure thus offers a probe of the nature and strength of the interlayer couplings.

Our theoretical analysis based on the model Eq.~\eqref{eq:H_interlayer} reveals distinct magnetization processes when field is aligned along different high symmetry directions. Within increasing field $\parallel [110]$, we find the angle between the N\'{e}el vectors of the two sublattices gradually increase from $\pi/2$ to $\pi$, at which point the system enters the collinear state. Meanwhile, the N\'{e}el vectors remain symmetric with respect to the field  [Fig.~\ref{Fig:Spinflop}(a)]. The onset field of the collinear state is given by $g\mu_B\mu_0 H_{c} = 16\sqrt{J_1KS^4}$. Note this process is a crossover as opposed to a phase transition in that no symmetry is spontaneously broken. 

By contrast, with the field $\parallel [100]$, the N\'{e}el vectors are initially pinned to the anticollinear state [Fig. \ref{Fig:Spinflop}(b)]. A spin flop transition occurs at $H_{c1}$, at which point the N\'{e}el vectors are no longer orthogonal and evolve toward the collinear state, whereby spontaneously breaking the $\pi$-rotation symmetry with respect to $[100]$. The system enters the collinear state at $H_{c2}$ although the collinear N\'{e}el vectors are not strictly orthogonal to the field. No symmetry breaking occurs at $H_{c2}$ and thus it constitutes a crossover. With increasing field, the collinear N\'{e}el vectors continuously approach the limit where they are orthogonal to the field. $H_{c1,c2}$ are determined by:
\begin{subequations}
\begin{align}
\frac{(g\mu_B \mu_0 H_{c1})^2}{32J_1} &= \sqrt{\mathcal{KD}},
\\
\frac{(g\mu_B \mu_0 H_{c2})^2}{32J_1} &= \sqrt{\frac{\mathcal{K}^2}{2}+\frac{\mathcal{K}}{2}\sqrt{\mathcal{K}^2+4\mathcal{D}^2}},
\end{align}
\end{subequations}
where $\mathcal{K} = 8KS^4$ and $\mathcal{D} = 12a^2DS^2 /(2a^2+c^2)$.

These predictions are confirmed experimentally by our dc magnetization measurements. The differential magnetization in $[110]$ direction shows a maximum near 5\,T, corresponding to the crossover from non-collinear to collinear states at $H_c$ \blue{[Fig.~\ref{Fig:Spinflop}(c)]}. By contrast, in the field $\parallel [100]$, we observe inflection points at $3.15(5)$\,T and $5$\,T [Fig.~\ref{Fig:Spinflop}(d)]. We identify the inflection near $3$\,T as the spin flop transition at $H_{c1}$ and the one near $5$\,T as the crossover at $H_{c2}$. This interpretation is further supported by ultrasound velocity measurements [Fig.~\ref{Fig:Spinflop}(f)]. When the field is applied in the $[100]$ direction, the relative speed of transverse sound wave shows a clear minimum at $3.27$\,T, indicative of a phase transition, but no anomaly is found at $5$\,T. Meanwhile, neutron diffraction measurement with field $\parallel [110]$ reveals a gradual increase (decrease) of magnetic Bragg peak intensities with even (odd) $L$ values up to the highest measured magnetic field of $4$\,T [Fig.~\ref{Fig:Spinflop}(e)], consistent with the picture of a gradual rotation of N\'{e}el vectors [Fig.~\ref{Fig:Spinflop}(a)].

Using the experimentally measured value of $H_{c1}$ and $H_{c2}$ in the $[100]$ direction, we estimate $DS^2 \approx 1.4\times 10^{-4}$\,meV and $KS^4 \approx 1.3\times10^{-4}$\,meV. Using these parameters, we determine the crossover field $\mu_0 H_{c} \approx 5$\,T in the $[110]$ direction, in agreement with the experiment. Meanwhile, the LSWT predicts all four branches of the spin waves are gapped. The interlayer interactions open two gaps with values $0.2$\,meV and $0.6$\,meV. The $0.6$\,meV gap is consistent with the observed spectral gap  $\Delta$[Fig.~\ref{Fig:INS}(d)], whereas the $0.2$\,meV gap is beyond the energy resolution of our measurements. 

We note that the alternative model for interlayer coupling, namely the dipolar coupling and the four-fold symmetric single-ion anisotropy, produces first-order spin flop transitions in the $[100]$ directions~\cite{Supplemental}, which is inconsistent with the experiment.

\textit{Discussion.} 
Having established the nature and strength of interlayer interactions [Eq.~\eqref{eq:H_interlayer}], we now discuss their microscopic origins. The dipolar coupling $D$ may originate from either the pseudo-dipolar coupling, commonly found in systems with strong spin-orbital coupling, or the magnetic dipolar interaction. Given the filled $t_{2g}$ shell of Cr$^{3+}$, we do not expect significant spin-orbital coupling and thus rules out the former possibility. Note that our case is very different from isostructural compounds with a second, magnetic rare earth sublattice, \eg $R_2$CuO$_4$ ($R$ = Ce, Pr, Nd)~\cite{Sa1993, Skanthakumar1993a, Sumarlin1994,  Li2005},  which could mediate the pseudo-dipolar coupling~\cite{Sachidanandam1997}. Instead, we find $D$ is naturally attributed to the magnetic dipolar coupling. Our magnetostatic calculation yields $DS^2 \approx 2\times10^{-4}$\,meV based on refined moment of $2.25\mu_\mathrm{B}$/Cr$^{3+}$, consistent with the estimate based on the spin flop field. Dipolar coupling is known to be crucial for rare-earth magnets with ice-like frustration~\cite{Hertog2000, Paddison2016, Dun2020} where the exchange interactions are small. Our work demonstrates that it can also play an important role in systems with comparatively much stronger exchange coupling.

The \emph{positive} biquadratic exchange interaction could be generated either by higher order virtual hopping processes in the superexchange~\cite{Hoffmann2020}, or more likely by quenched disorder due to the La/Sr mixing through the ObD mechanism~\cite{Henley1989, Smirnov2017}. We also note that the combination of dipolar interaction and a \emph{negative} biquadratic exchange, produced by the thermal or quantum ObD, would stabilize a collinear order with the spins in the [110] direction, which may explain the 3D ordering in Sr$_2$CuO$_2$Cl$_2$ or LaSrFeO$_4$. This observation motivates further investigation of quenched disorder to control magnetic order in frustrated magnets or spintronic devices. 

The experimental observation of the anticollinear order in LSCrO uncovers a new territory in the phase diagram of the AB-stacked square-lattice antiferromagnet. In contrast to the collinear magnetic states displayed by all related materials, the anticollinear order in LSCrO exhibits a rich and unique magnetic field evolution stemming from interlayer effects that are merely a few parts-per-million of the main exchange interaction. A systematic study of the temperature-field phase diagram of LSCrO and its materials relatives is poised to reveal more surprises in this canonical family of geometrically frustrated magnets.

\begin{acknowledgements}
We thank Cristian Batista and Hitesh Changlani for helpful discussions. This research used resources at the High Flux Isotope Reactor and Spallation Neutron Source, a DOE Office of Science User Facility operated by the Oak Ridge National Laboratory. The work at Institute of Physics was supported by the National Natural Science Foundation of China (Grant No.~11974396, 12025408, 11874400, 12188101), the Ministry of Science and Technology (2018YFA0305700), and the Strategic Priority Research Program of the Chinese Academy of Sciences (Grant No.~XDB33020300). The work of Q.H. and H.Z. at the University of Tennessee was supported by the National Science Foundation under award DMR-2003117. The work of Z.L.D., X.B. and M.M. at Georgia Tech (neutron scattering experiment and data analysis)  was supported by the U.S. Department of Energy, Office of Science, Basic Energy Sciences, Materials Sciences and Engineering Division under award DE-SC-0018660.
\end{acknowledgements}

\bibliography{LaSrCrO4}

\newpage
\setcounter{figure}{0}
\renewcommand{\thefigure}{S\arabic{figure}}
\renewcommand{\theequation}{S\arabic{equation}}

\onecolumngrid
\begin{center}
{\huge Supplemental Material}
\end{center}

\tableofcontents

\section{Experimental Methods}
\subsection{crystal growth} 
\begin{figure}[tbp!]
	\linespread{1}
	\par
	\begin{center}
		\includegraphics[width= 5 in ]{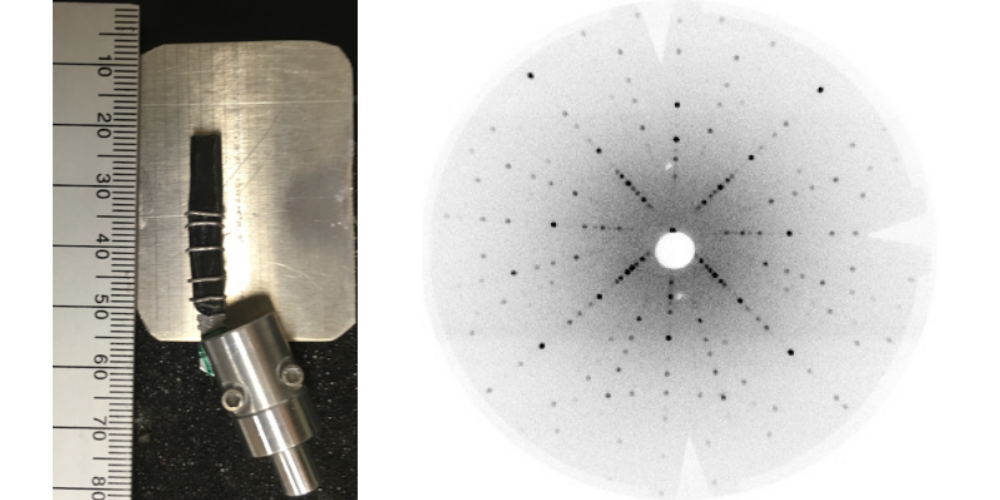}
	\end{center}
	\par
	\caption{\label{Fig:Crystal} Left: The single crystal used for the inelastic neutron scattering measurements at SEQUOIA. Right: Laue diffraction pattern along the [001] direction of a single crystal LaSrCrO$_4$.}
\end{figure}

Single Crystals of LaSrCrO$_4$ were synthesized using a floating zone technique. In order to grow the single crystals, powder sample was first synthesized from a stoichiometric mixture of La$_2$O$_3$ (baked before usage), SrCO$_3$, and Cr$_2$O$_3$ under flowing 10$\%$H$_2$/Ar atmosphere at 1350 $^\circ$C for 40h with several intermediate re-grindings. 
The synthesized powder was pressed into cylindrical rods (of approximately 6 mm in diameter and 70–80 mm in length) and then melted at higher temperatures in a two-mirror optical floating zone furnace (NEC, Conan Inc., equipped with two 1500W halogen lamps) under 1 atmosphere gas of 4$\%$H$_2$/Ar. 
The best crystal was obtained using a growth voltage of 92 V and a pulling rate of 35 mm/h. Slight evaporation of melted mixture was observed during growth. The obtained single crystal was black in color and naturally cleaves into shining surfaces that are perpendicular to the crystallographic c-axis [Fig. \ref{Fig:Crystal} left], reflecting the quasi-2D nature of the layered perovskite structure.
The crystals were oriented by Laue back diffraction for subsequently measurements [Fig. \ref{Fig:Crystal} right].

\subsection{Magnetic measurements} 
\begin{figure}[tbp!]
	\linespread{1}
	\par
	\begin{center}
		\includegraphics[width= 5 in ]{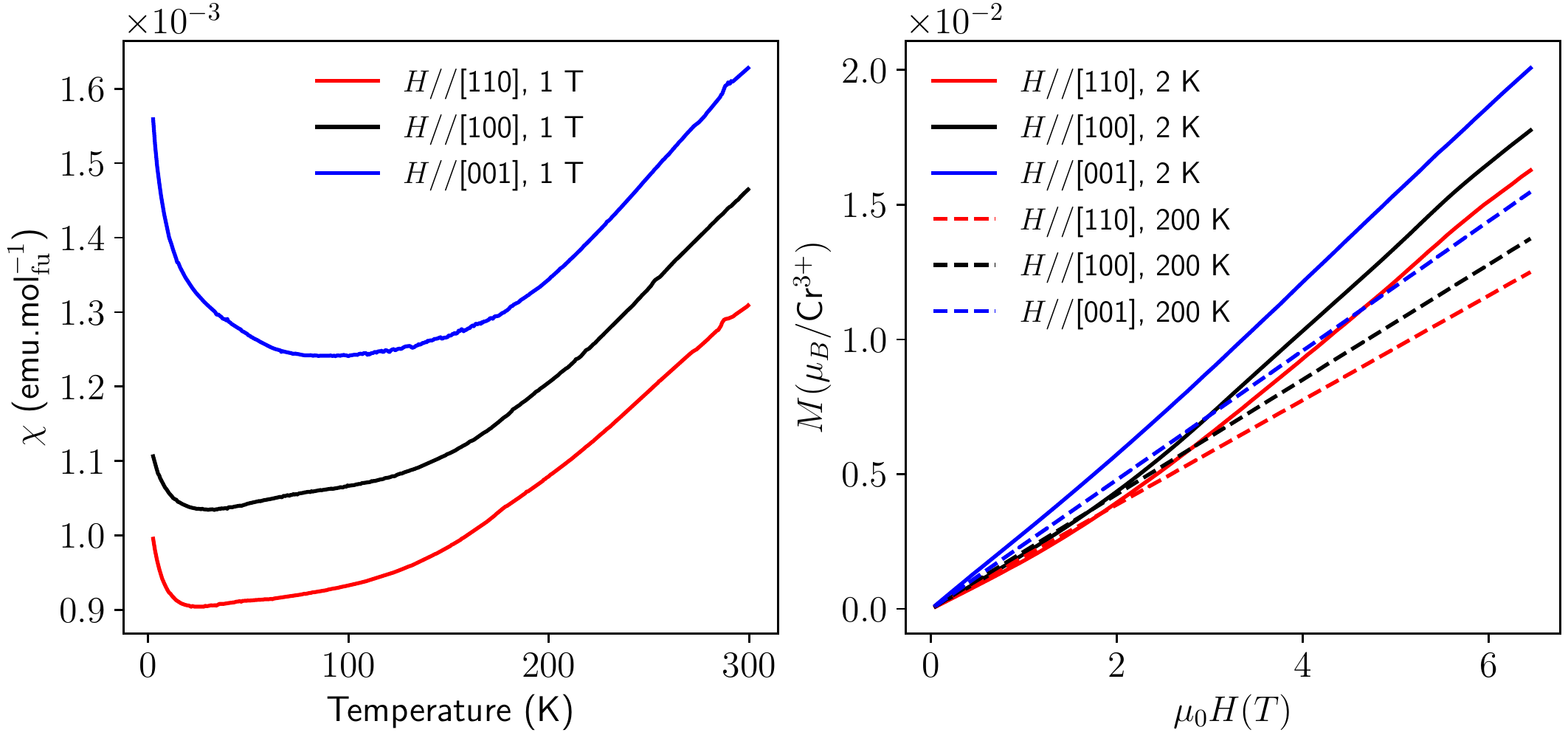}
	\end{center}
	\par
	\caption{\label{Fig:VSM} Left: dc susceptibility under a magnetic field of 1 T that is applied along three high-symmetry directions. The increases of $\chi$ at low temperature are due to Curie tails from magnetic impurities. Right: magnetization curves measured at 2 K and 200 K with field applied along three high-symmetry directions.}
\end{figure}

Magnetic susceptibility measurements were made using a Quantum Design Magnetic Properties Measurement System with a superconducting interference device (SQUID) magnetometer. Measurements were made after cooling in zero field of $\mu_0$H= 1 T over the temperature range 2 K to 300 K [Fig. \ref{Fig:VSM} left]. A weak slope change was observed at the N\'eel temperature of 170 K.  Isothermal magnetization $M(H)$ measurements were made using the same SQUID at  temperatures of 2 K and 200 K between 0 and 6.5 T [Fig. \ref{Fig:VSM} right]. 

\subsection{Neutron-scattering measurements} 
Single crystal neutron diffraction measurements were carried out using the Four-Circle Diffractometer (HB3A) \cite{Chakoumakos2011} at the High Flux Isotope Reactor, and diffuse scattering measurements using the Elastic Diffuse Scattering Spectrometer (CORELLI) \cite{ye2018}  at the Spallation Neutron Source, both located at Oak Ridge National Laboratory. For the HB3A measurements, a small single crystal was cooled down to a base temperature of 4 K,  and measured using  a constant neutron wavelength of $\lambda = 1.003$~\AA.  For the CORELLI measurements, a single crystal was  oriented in the (HHL) scattering plane and cooled down to 2\,K using a orange cryostat inside a 5\,T superconducting magnet. 

Inelastic neutron scattering measurements were carried out using the Fine-Resolution Fermi Chopper Spectrometer (SEQUOIA) \cite{Granroth2010} at the Spallation Neutron Source, Oak Ridge National Laboratory.  A single crystal $\sim$3\,g (size: 4 mm$\times$40 mm$\times$6 mm)  [Fig. \ref{Fig:Crystal} left] was oriented in the (HHL) scattering plane cooled down to 5\,K with a closed-cycle refrigerator. Magnon excitations were mapped out with incident neutron energies of $120$, $20$, and $8$ meV with sample rotation of 2$^\circ$/step. Constant energy cut of the 120 meV dataset in the [HK0] scattering plane was summarized in Fig.~\ref{Fig:120meV}. The data was integrated along L direction and symmetrized according to the point group symmetry of the Cr site, resulting in the dispersion curve shown in the main text. 

\begin{figure}[tbp!]
	\linespread{1}
	\par
	\begin{center}
		\includegraphics[width= \columnwidth]{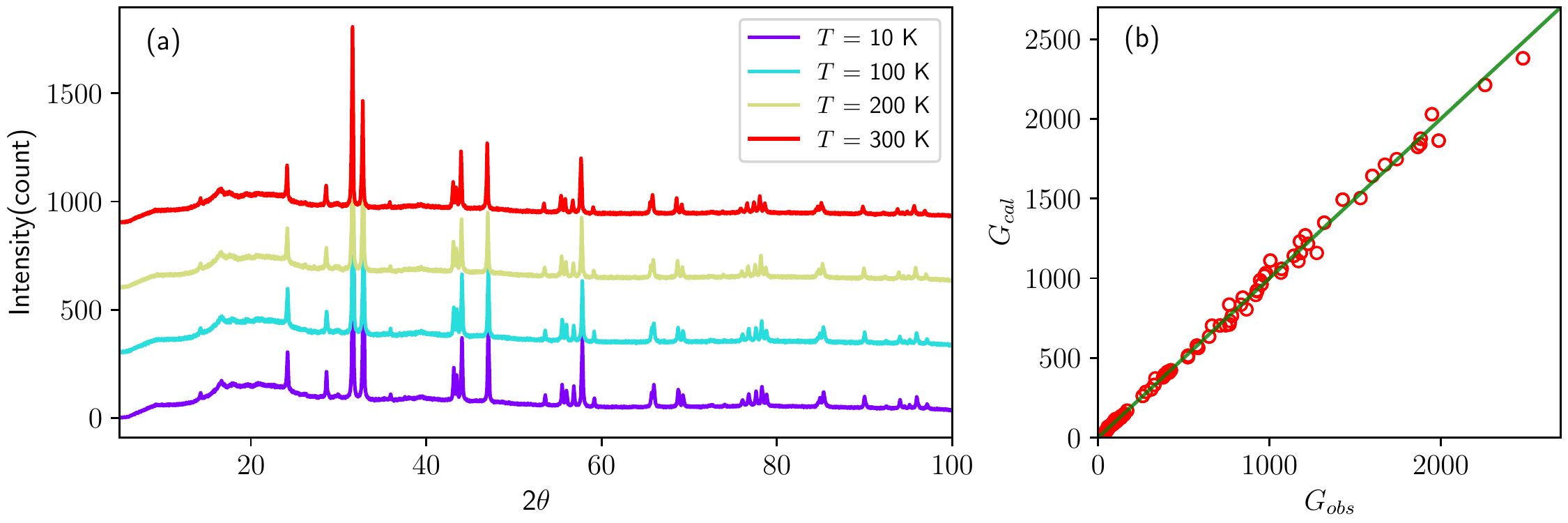}
	\end{center}
	\par
	\caption{\label{Fig:XRD} (a) Powder XRD diffraction patterns of LaSrCrO$_4$ at various temperatures between 300 K to 10 K, showing that there is no structural phase transitions.  (b) Structural refinements of the nuclear Bragg peaks measured using HB3A Four-circle diffractometer at 4 K. $G_{obs}$ and $G_{calc}$ represent observed and calculated intensities, respectively.}
\end{figure}

\subsection{Refinements of nuclear and magnetic structures }
Rietveld refinements of the nuclear and magnetic structures are done using the 4 K neutron diffraction data measured at HB3a. Nuclear reflections are collected for Bragg peaks at wave-vectors $\mathbf{Q} = (H,K,L)$, and magnetic reflections are collected at $\mathbf{Q} = (H+1/2,K+1/2,L+1/2)$, where $H, K, L$ are integer numbers.  

Crystal structural refinement was carried out using the FULLPROF suite of programs. The observed nuclear Bragg peaks intensities agrees well with layered perovskite structure of the I4/mmm space group. Refined crystallographic parameters are given in TABLE I.  

Magnetic structural analysis was carried out using the Bilbao Crystallographic Server \cite{Perez-Mato2015a} using two propagation vectors, $k_1 = (0.5, 0.5, 0)$ and $k_2 = (0.5, -0.5, 0)$.  The maximal magnetic space groups for $I4/mmm$ space group and the two propagation vectors are $P_C4_2/ncm$, $P_C4_2/mbm$, and $P_C4_2/nbm$.  $P_C4_2/ncm$ corresponds to the observed spin order in LaSrCrO$_4$. $P_C4_2/mbm$ corresponds a different anticollinear state, i.e., $\phi_a = 0$, $\phi_b = -\pi/2$ shown in Fig.1(e) of the main text. $P_C4_2/nbm$ describe a N\'{e}el order with easy axis anisotropy, i.e.  N\'{e}el order vectors are pointing along crystallographic $c$-axis.

\begin{table*}
\caption{ Crystallographic parameters and selected bond lengths for LaSrCrO$_{4}$ for single crystal refinement of HB3a neutron diffraction data at 4 K.}
\begin{tabular}{cccccccc}
\hline
&\multicolumn{7}{c}{Space group: I4/mmm. $T$ = 4\,K}\\
\hline
Atom& $x$ & $y$ & $z$ & $u_{11}$  ($\mathrm{\AA^2}$)  & $u_{22}$  ($\mathrm{\AA^2}$)  & $u_{33}$  ($\mathrm{\AA^2}$) & Occu. \\
    \hline
 La & 0 & 0 & 0.35957(6)   & 0.0012(4) & 0.0012(4) & 0.0006(1)  & 0.5  \\
 Sr & 0 & 0 & 0.35957(6)   & 0.0012(4) & 0.0012(4) & 0.0006(1)  & 0.5  \\
 Cr & 0 & 0 & 0            & 0.0009(9) & 0.0009(9) & 0.0005(1)  & 1  \\
 O1 & 0.5 & 0 & 0   &        0.0043(7) & 0.0043(7) & 0.0009(1)  & 1.00(1)    \\
 O2 & 0 & 0 & 0.16553(8)   & 0.0128(4) & 0.0128(4) & 0.0009(1)  & 0.99(1)  \\
\cline{1-8}
  &\multicolumn{7}{c}{\emph{a}=\emph{b}=3.853(6) $\mathrm{\AA}$,~\emph{c}=12.475(4) $\mathrm{\AA}$} \\
    &\multicolumn{7}{c}{Cr-O(1)=2.065(1) $\mathrm{\AA}$}\\
   &\multicolumn{7}{c}{Cr-O(2)=1.9265(8) $\mathrm{\AA}$}\\
  &\multicolumn{7}{c}{R$_{w}$ = 4.6\%, R$_{f}$ = 2.7\%, $\chi^2$ = 2.51\%} \\
\hline
\end{tabular}
\end{table*}

\subsection{Ultrasonic measurement}
All measurements have been obtained using a single crystal with parallel faces normal to the crystallographic $a$-axis. 30 MHz LiNbO$_3$ transducers were mounted on those faces in order to realized sound velocity measurements in the transmission configuration. A sample of 4.08 mm in length, along the direction of propagation (x $\parallel$ $a$-axis), was necessary in order to determine the velocity of longitudinal modes $V_{Lx}$ and transverse modes with a polarization along (y $\parallel$ $b$-axis), $V_{TxPy}$. The data, realized at 90 MHz using a pulsed acoustic interferometer, were used to explore the spin flop transitions of LaSrCrO$_4$ with the field parallel the a and c-axis up to 14 T between 2 K and 200 K. 

\begin{figure}[tbp!]
	\linespread{1}
	\par
	\begin{center}
		\includegraphics[width= \columnwidth ]{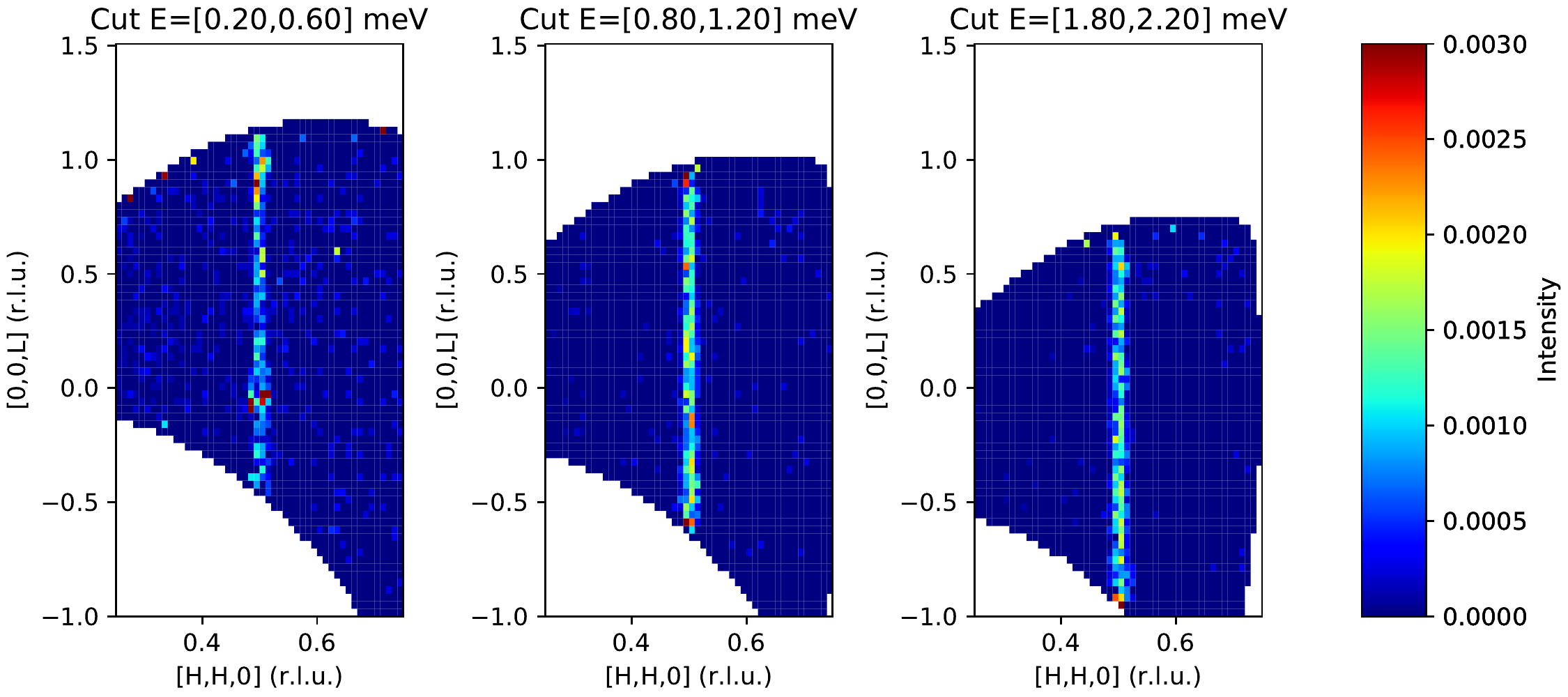}
	\end{center}
	\par
	\caption{\label{Fig:8meVl} Constant energy slices through the quasielastic scattering in the (HHL) plane at different neutron energy transfer measured on SEQUOIA at $T$ = 5 K, showing that the magnon excitations is dispersiveless along L direction.  Data was measured with neutron incident energy of 8 meV and integrated for $\mathrm{[K\bar{K}0]}$ direction within K = [-0.02, 0.02] reciprocal lattice unit (r.l.u.). }
\end{figure}

\begin{figure}[tbp!]
	\linespread{1}
	\par
	\begin{center}
		\includegraphics[width= \columnwidth ]{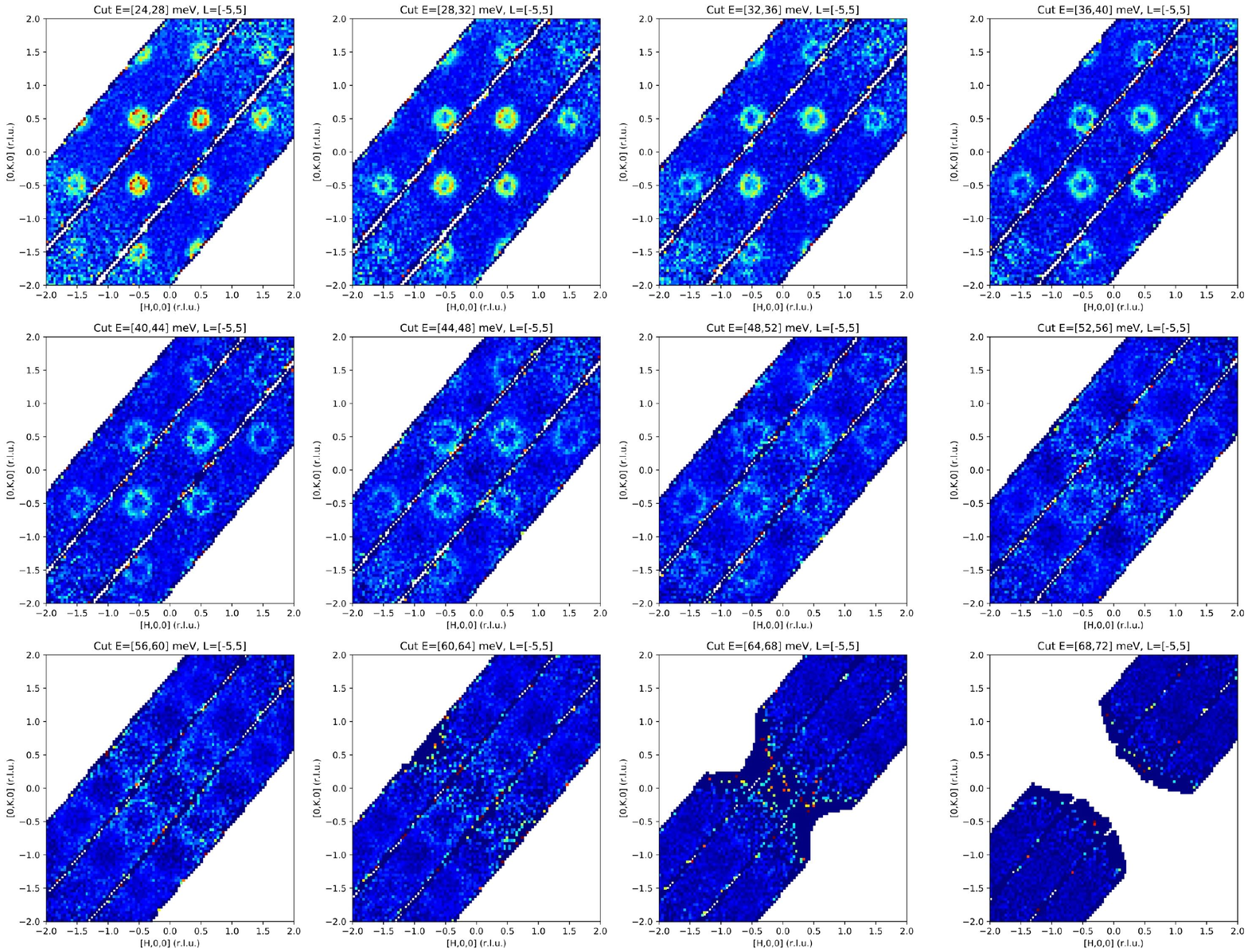}
	\end{center}
	\par
	\caption{\label{Fig:120meV} Constant energy slices through the inelastic scattering in  the (HK0) scattering plane at different neutron energy transfer measured on SEQUOIA at $T$ = 5 K.  Data was measured with neutron incident energy of 120 meV and integrated along L direction within L = [-5, 5] reciprocal lattice unit  (r.l.u.)}
\end{figure}

\section{Estimate the energy scale of interlayer coupling}
The 3D N\'eel ordering temperature is relatively small compared to the intralayer exchange interaction in LaSrCrO$_4$, i.e., $k_BT_N/[J_1 S(S+1)] = 0.391$. The N\'eel temperature of a quasi-two-dimensional spin-$S$ Heisenberg antiferromagnet is given by~\cite{Majlis1992, Yasuda2005}: 
\begin{equation} \label{eq:interlayer}
\frac{k_B T_\mathrm{N}}{J_1 S(S+1)}= \frac{4 \pi \rho_{\mathrm{s}}} {b-\ln \left(J' / J_1\right)},
\end{equation}
where $\rho_s$ is the spin stiffness, $b$ is a dimensionless constant, and $J'$ is the interlayer coupling. As $\rho_s$ and $b$ are unknown for $S=3/2$, we utilize the published N\'{e}el temperature data for $S=1$ and $S=\infty$~\cite{Yasuda2005}, obtained by quantum and classical Monte Carlo simulations, respectively to constrain the value of $J'/J$. Fitting the $T_\mathrm{N}$ data for $S=1$ to Eq.~\eqref{eq:interlayer} yields $\rho_{\mathrm{s}} = 0.68/S/(S+1) =0.34$ and $b=3.12$. Using the value of $k_BT_N/[J_1 S(S+1)]$ for LaSrCrO$_4$ and solving for $J'$, we obtain $J^{\prime}/J_1 = 4.1\times10^{-4}$. On the other hand, fitting the $T_\mathrm{N}$ data for $S=\infty$ yields $\rho_s = 0.84$ and $b = 10.9$, from which we estimate $J/J_1 = 1.1\times10^{-7}$ in LaSrCrO$_4$. These two estimates could be viewed as the upper and lower bounds on $J'/J$ in LaSrCrO$_4$. As $J\sim 10$ meV in LaSrCrO$_4$, we deduce that the order of magnitude of the interlayer coupling is between $10^{-3}$ meV and $10^{-6}$ meV, presumably closer to the upper limit.

\section{Symmetry analysis for interlayer coupling}

The magnetic order of LaSrCrO$_4$ comprises of two sublattices, each hosting a three-dimensional N\'{e}el order. In this section, we perform the symmetry analysis on the coupling between the two N\'{e}el vectors .

Let $\phi_a$ ($\phi_b$) be the azimuthal angle of the N\'{e}el vector of the A (B) sublattice. For the A sublattice (crystallographic site label $(i,j,k)$), we define $\phi_a$ to be the angle between the spin on site $(0,0,0)$ and the crystallographic $a$ axis. For the B sublattice (crystallographic site label $(i+\frac{1}{2},j+\frac{1}{2},k+\frac{1}{2})$), we define $\phi_b$ to be the angle between the spin on site $(\frac{1}{2},\frac{1}{2},\frac{1}{2})$ and the crystallographic $a$ axis. Generically, the coupling between the two Neel vectors can be expanded as Fourier series:
\begin{align}
E(\phi_a,\phi_b) = \sum_{m,n} C_{m,n} \cos(m\phi_a+n\phi_b) + S_{m,n}\sin(m\phi_a+n\phi_b),
\end{align} 
where $m,n$ run over all integers. $C_{m,n}$ and $S_{m,n}$ are real coefficients. 

We use symmetry to constrains the possible form of $E(\phi_a,\phi_b)$. To this end, we use the fact that the space group $I4/mmm$ is symmorphic; it is therefore sufficient to consider the operations of the point group $D_{4h}$. The point group $D_{4h}$ is generated a 4-fold rotation with respect to the crystallographic $c$ axis ($C_4$), a mirror operation whose norm is along the crystallographic $a$ axis ($m$), and an spatial inversion $i$. Without loss of generality, we take the 4-fold axis and the mirror plane pass through an A site. We find:
\begin{align}
(\phi_a,\phi_b) \overset{C_4}{\to} (\phi_a+\frac{\pi}{2},\phi_b-\frac{\pi}{2});
\quad
(\phi_a,\phi_b) \overset{m}{\to} (-\phi_a,\pi-\phi_b).
\end{align}
Meanwhile, under the time reversal operation:
\begin{align}
(\phi_a,\phi_b) \overset{\mathcal{T}}{\to} (\phi_a+\pi,\phi_b+\pi).
\end{align}
Finally, the inversion exchanges $\phi_a$ and $\phi_b$. As a result, the interaction must be symmetric with respect to $\phi_{a,b}$:
\begin{align}
E(\phi_a,\phi_b) = E(\phi_b,\phi_a).
\end{align}
These symmetries are sufficient to constrain the interactions; adding more symmetries do not yield more independent constrains.

The exchange symmetry between A and B sublattices constrains the harmonics must come in pairs:
\begin{align}
\cos(m\phi_a+n\phi_b)+\cos(n\phi_a+m\phi_b),\,
\sin(m\phi_a+n\phi_b)+\sin(n\phi_a+m\phi_b).
\end{align}
Under the mirror operation, these terms become:
\begin{align}
(-)^n \cos(m\phi_a+n\phi_b)+(-)^m \cos(n\phi_a+m\phi_b),\,
(-)^{n+1} \sin(m\phi_a+n\phi_b)+(-)^{m+1}\sin(n\phi_a+m\phi_b).
\end{align}
These conditions require that, for the $\cos$ harmonics, both $m$ and $n$ must be \emph{even}; whereas, for the $\sin$ harmonics, both $m$ and $n$ must be \emph{odd}. These conditions automatically enforce the time reversal symmetry.

The remaining task is to enforce the $C_4$ symmetry. Note the symmetry operations cannot mix harmonics of different orders $(m,n)$. Meanwhile, the parity of $(m,n)$ are distinct for the $\cos$ terms and $\sin$ terms. It follows that $\cos$ terms and $\sin$ terms cannot mix under symmetry operations. Said differently, the $\cos$ terms must transform to itself, and so do the $\sin$ terms. On the other hand, under the action of $C_4$:
\begin{align}
\cos(m\phi_a+n\phi_b)+\cos(n\phi_a+m\phi_b) \to (-1)^{\frac{m-n}{2}} [\cos(m\phi_a+n\phi_b)+\cos(n\phi_a+m\phi_b)],
\nonumber\\
\sin(m\phi_a+n\phi_b)+\sin(n\phi_a+m\phi_b) \to (-1)^{\frac{m-n}{2}} [\sin(m\phi_a+n\phi_b)+\sin(n\phi_a+m\phi_b)].
\end{align}
Here, we have used the fact that $m-n$ is always an even number (and hence $(m-n)/2$ is an integer). The invariance requires that $m-n$ must be multiples of 4.

We are now ready to write down all the symmetry allowed interactions by their order of harmonics $p = |m|+|n|$. Terms with odd $p$ are forbidden by symmetry.  At $p=2$, there is only 1 \emph{linearly} independent, symmetry allowed term:
\begin{align}
\sin(\phi_a+\phi_b).
\end{align}
This term can be generated by the magnetic dipole coupling or the pseudo-dipolar coupling between the two sublattices. At $p=4$, there are 3 \emph{linearly} independent, symmetry allowed terms:
\begin{align}
\cos(4\phi_a)+\cos(4\phi_b),\quad
\cos(2\phi_a-2\phi_b),\quad
\sin(3\phi_a-\phi_b)+\sin(3\phi_a-\phi_b),\quad
\cos(2\phi_a+2\phi_b).
\end{align}
The first term can be generated by an in-plane single-spin crystalline anisotropic term. The second term can be interpreted as the biquadratic exchange interaction. The third term does not have any obvious microscopic interpretations; however, it can be viewed as a product of the $p=1$ term with the $\cos(4\phi_a)+\cos(4\phi_b)$ term. Likewise, the last term can be viewed as the square of the $p=1$ term. Thus, the last two terms are not \emph{algebraically} independent.

Keeping \emph{algebraically independent} terms up to $p=4$, we arrive at the two minimal models for the interlayer coupling. The first model, which we dub the dipole-biqaudratic model, is given by:
\begin{align}
E(\phi_a,\phi_b) = S_{1,1}\sin(\phi_a+\phi_b) + C_{2,-2}\cos(2\phi_a-2\phi_b).
\end{align}
Stabilizing the anticollinear state in LaSrCrO$_4$ require $S_{1,1}<0$ and $C_{2,-2}>0$. Minimizing the energy yields four symmetry-related solutions: $\phi_a = 0$, $\phi_b = \pi/2$; $\phi_a = \pi$, $\phi_b = -\pi/2$; $\phi_a = \pi/2$, $\phi_b = 0$; and $\phi_a = -\pi/2$, $\phi_b = \pi$. Microscopically, the dipole-biquadratic model consists of both dipole and biquadratic exchange interaction between spins between the third neighbor pairs:
\begin{align}
H' = \sum_{\langle ij\rangle_3} D(\mathbf{S}_i\cdot\mathbf{S}_j-3(\mathbf{S}_i\cdot\hat{n}_{ij})(\mathbf{S}_j\cdot\hat{n}_{ij}) ) + K(\mathbf{S}_i\cdot\mathbf{S}_j)^2.
\end{align}
Here, the summation is over all third neighbor bonds. $\hat{n}_{ij}$ is the unit vector pointing from site $i$ to site $j$. $D>0$ and $K>0$ are model parameters.

The second model, which we dub the dipole-anisotropy model, is given by:
\begin{align}
E(\phi_a,\phi_b) = S_{1,1}\sin(\phi_a+\phi_b) + C_{4,0}(\cos(4\phi_a)+\cos(4\phi_b)).
\end{align}
Stabilizing the anticollinear state found in LaSrCrO$_4$ require $S_{1,1}<0$ and $C_{4,0}<0$. Microscopically, this model consists of dipole interaction between spins on adjacent layers, and a single-ion anisotropy term:
\begin{align}
H' = \sum_{\langle ij\rangle_3} D(\mathbf{S}_i\cdot\mathbf{S}_j-3(\mathbf{S}_i\cdot\hat{n}_{ij})(\mathbf{S}_j\cdot\hat{n}_{ij}) ) + \frac{A'}{2} \sum_i ((S^x_i)^2 (S^y_i)^2+(S^y_i)^2 (S^x_i)^2).
\end{align}
Here, $D>0$ and $A'>0$ are model parameters.

\section{Spin flop transitions of the dipole-biquadratic model}

In this section, we analyze the spin flop transitions of the dipole-biquadratic model. We determine analytically the critical field and the magnetization curve. These analytic results allow us to estimate the strength of the effective dipole coupling and the biquadratic coupling. In our calculations, we assume that the nearest neighbor Heisenberg exchange interaction $J_1$ is significantly larger than the Zeeman energy as well as all the other spin interactions. 

\subsection{Field along the $[100]$ direction}

We first consider the spin flop transitions when the field is applied along the $[100]$ direction. In the absence of magnetic field, the spins belonging to the even layers (dubbed A sublattice) form a three-dimensional  N\'{e}el order, whereas thee spins belonging to the odd layers (dubbed B sublattice) form another three-dimensional  N\'{e}el order. Since the Zeeman energy is significantly weaker than the nearest neighbor exchange interaction, the N\'{e}el orders are only slightly distorted. Let the $\phi_a$ and $\phi_b$ be the azimuthal angle of the Neel vectors in the A and B sublattices, respectively. Here, $\phi_a$ ($\phi_b$) is defined as the azimuthal angle of the spin on site $(0,0,0)$ ($(1/2,1/2,1/2)$).  The energy of these two subsystems in the presence of the magnetic field is given by:
\begin{align}
\frac{E}{N} = -\frac{(g\mu_B B)^2}{32J_1}(\sin^2\phi_a+\sin^2\phi_b),
\end{align}
where $N$ is the total number of spins, $g$ the Land\'{e} g-factor, $\mu_0$ the Bohr magneton. This energy alone favors both N\'{e}el vectors to be orthogonal with the field; it must compete with the dipole/biquadratic coupling between the two subsystems,
\begin{align}
\frac{E'}{N} = 4KS^4\cos^2(\phi_a-\phi_b) - \frac{12a^2}{2a^2+c^2}DS^2\sin(\phi_a+\phi_b).
\end{align}
Here, $a$ and $c$ are the lattice constants. The magnetic ground state is determined by minimizing the total energy:
\begin{align}
E_\mathrm{tot} = E+E'.
\end{align}

It is convenient to define a new pair of variables:
\begin{align}
\phi_+ = \phi_a+\phi_b;\quad \phi_- = \phi_a-\phi_b.
\end{align}
We rewrite the total energy as,
\begin{align}
\frac{E_\mathrm{tot}}{N} = \mathcal{B}\cos\phi_+\cos\phi_- + \frac{\mathcal{K}}{2}\cos^2\phi_- - \mathcal{D}\sin\phi_+.
\end{align}
Here, we have defined a set of parameters for the sake of brevity:
\begin{align}
\mathcal{B} = \frac{(g\mu_B B)^2}{32J_1};
\quad
\mathcal{K} = 8KS^4;
\quad
\mathcal{D} = \frac{12a^2}{2a^2+c^2}DS^2.
\end{align}
The stationary condition is thus given by:
\begin{align}
\sin\phi_- (\mathcal{B}\cos\phi_+ + \mathcal{K}\cos\phi_-) = 0;
\quad
\mathcal{B}\sin\phi_+\cos\phi_- + \mathcal{D}\cos\phi_+ = 0.
\end{align}
The Hessian matrix is given by:
\begin{align}
M = \begin{bmatrix}
-\mathcal{B}\cos\phi_+\cos\phi_- -\mathcal{K}\cos2\phi_- & \mathcal{B}\sin\phi_+\sin\phi_- \\
\mathcal{B}\sin\phi_+\sin\phi_- & -\mathcal{B}\cos\phi_+\cos\phi_- + \mathcal{D}\sin\phi_+
\end{bmatrix}.
\end{align}

We find three solutions to the stationary condition. They are:
\begin{itemize}

\item \textbf{Orthogonal state}. This state is identical to the zero-field magnetic ground state. It is corresponds to:
\begin{align}
\phi_+ = \frac{\pi}{2};\quad\phi_- = \pm\frac{\pi}{2}.
\end{align}
The Hessian matrix reads:
\begin{align}
M = \begin{bmatrix}
\mathcal{K} & \pm\mathcal{B} \\
\pm\mathcal{B} & \mathcal{D}
\end{bmatrix}.
\end{align}
The stability of this state requires
\begin{align}
\mathcal{B}^2 \le \mathcal{B}^2_{c1} = \mathcal{KD}.
\end{align}

\item \textbf{Collinear state}. In this state, the two N\'{e}el vectors are collinear. They correspond to the following two symmetry related solutions:
\begin{subequations}
\begin{align}
\phi_- = 0,\quad\phi_+ = \pi-\arcsin\frac{\mathcal{D}}{\sqrt{\mathcal{B}^2+\mathcal{D}^2}}.
\end{align} 
or
\begin{align}
\phi_- = \pi,\quad \phi_+ = \arcsin\frac{\mathcal{D}}{\sqrt{\mathcal{B}^2+\mathcal{D}^2}}.
\end{align}
\end{subequations}
The Hessian matrix reads:
\begin{align}
M = \begin{bmatrix}
\frac{\mathcal{B}^2}{\sqrt{\mathcal{B}^2+\mathcal{D}^2}}-\mathcal{K} & 0 \\
0 & \frac{\mathcal{B}^2}{\sqrt{\mathcal{B}^2+\mathcal{D}^2}}
\end{bmatrix}.
\end{align}
The stability of this state requires
\begin{align}
\mathcal{B}^2>\mathcal{B}^2_{c2} = \frac{\mathcal{K}^2}{2} (1+\sqrt{1+\frac{4\mathcal{D}^2}{\mathcal{K}^2}}).
\end{align}

\item \textbf{Intermediate state}. In this state, the N\'{e}el vectors are neither orthogonal nor collinear. It corresponds to the solutions:
\begin{subequations}
\begin{align}
\phi_+ = \arcsin\frac{\mathcal{KD}}{\mathcal{B}^2},
\quad
\phi_- = \pm(\pi-\arccos(\frac{\mathcal{B}}{\mathcal{K}}\sqrt{1-\frac{\mathcal{K}^2\mathcal{D}^2}{\mathcal{B}^4}})),
\end{align}
and
\begin{align}
\phi_+ = -\arcsin\frac{\mathcal{KD}}{\mathcal{B}^2},
\quad
\phi_- = \pm\arccos(\frac{\mathcal{B}}{\mathcal{K}}\sqrt{1-\frac{\mathcal{K}^2\mathcal{D}^2}{\mathcal{B}^4}}).
\end{align}
\end{subequations}
Apparently, this state exists if and only if:
\begin{align}
\mathcal{B}^2_{c1}\le \mathcal{B}^2 \le \mathcal{B}^2_{c2}.
\end{align}
\end{itemize}

To conclude, at low magnetic field ($\mathcal{B}<\mathcal{B}_{c1}$), the system is in the orthogonal state; at intermediate field ($\mathcal{B}_{c2}<\mathcal{B}<\mathcal{B}_{c1}$), the system is in the intermediate field; at high field  ($\mathcal{B}_{c2}<\mathcal{B}$), the system is in the collinear state. 

We now compute the magnetization curve. The magnetization parallel ($M^\parallel$) and orthogonal ($M^\perp$) to the magnetic field is given by:
\begin{align}
\frac{M^\parallel}{N} = \frac{(g\mu_B)^2B}{16J_1}(1-\cos\phi_+\cos\phi_-);
\quad
\frac{M^\perp}{N} = -\frac{(g\mu_B)^2B}{16J_1}\sin\phi_+\cos\phi_-.
\end{align}
Using the previously obtained solutions for $\phi_\pm$, we find:
\begin{subequations}
\begin{align}
\frac{M^\parallel}{N} &= \frac{(g\mu_B)^2B}{16J_1}\times\left\{
\begin{array}{cc}
1 & (\mathcal{B}\le \mathcal{B}_{c1}) \\
1+\frac{\mathcal{B}}{\mathcal{K}}-\frac{\mathcal{K}\mathcal{D}^2}{\mathcal{B}^3} & (\mathcal{B}_{c1} \le \mathcal{B} \le \mathcal{B}_{c2}) \\
1+\frac{\mathcal{B}}{\sqrt{\mathcal{B}^2+\mathcal{D}^2}} & (\mathcal{B}_{c2} \le \mathcal{B})
\end{array}\right. ;
\\
\frac{M^\perp}{N} &= \frac{(g\mu_B)^2B}{16J_1}\times\left\{
\begin{array}{cc}
0 & (\mathcal{B}\le \mathcal{B}_{c1}) \\
\pm\frac{\mathcal{D}}{\mathcal{B}}\sqrt{1-\frac{\mathcal{K}^2\mathcal{D}^2}{\mathcal{B}^4}} & (\mathcal{B}_{c1} \le \mathcal{B} \le \mathcal{B}_{c2}) \\
\pm\frac{\mathcal{D}}{\sqrt{\mathcal{B}^2+\mathcal{D}^2}} & (\mathcal{B}_{c2} \le \mathcal{B})
\end{array}\right. 
\end{align}
\end{subequations}
In the expression for $M^\perp$, the plus and minus signs correspond to the two symmetry-related magnetic ground states.

\subsection{Field along the $[110]$ direction}

When the field is applied along the $[110]$ direction, the total energy now reads:
\begin{align}
\frac{E_\mathrm{tot}}{N} = \mathcal{B}\sin\phi_+\cos\phi_- + \frac{\mathcal{K}}{2}\cos^2\phi_- -\mathcal{D}\sin\phi_+.
\end{align}
Here, the definition of $\phi_\pm$, $\mathcal{B}$, $\mathcal{K}$, and $\mathcal{D}$ are the same as before. In comparison with the energy in $[110]$ field, the energy due to the magnetic field (the first term) changes its form, whereas the other two terms stay the same. The stationary condition reads:
\begin{align}
\sin\phi_- (\mathcal{B}\sin\phi_+ + \mathcal{K}\cos\phi_-) = 0;
\quad
\cos\phi_+ (\mathcal{B}\cos\phi_- - \mathcal{D}) = 0.
\end{align}
The Hessian matrix reads:
\begin{align}
M = \begin{bmatrix}
-\mathcal{B}\sin\phi_+\cos\phi_- - \mathcal{K}\cos2\phi_- & -\mathcal{B}\cos\phi_+ \sin\phi_- \\
-\mathcal{B}\cos\phi_+ \sin\phi_- & -\mathcal{B}\sin\phi_+\cos\phi_- + \mathcal{D}\sin\phi_+
\end{bmatrix}.
\end{align}

We find three solutions to the stationary condition:
\begin{itemize}
\item \textbf{Noncollinear state}. This state is characterized by:
\begin{align}
\phi_+ = \frac{\pi}{2} ;
\quad
\phi_- = \pm(\pi-\arccos(\frac{\mathcal{B}}{\mathcal{K}})).
\end{align}
The Hessian matrix reads:
\begin{align}
M = \begin{bmatrix}
\mathcal{K}-\frac{\mathcal{B}^2}{\mathcal{K}} & 0\\
0 & \frac{\mathcal{B}^2}{\mathcal{K}}+\mathcal{D}
\end{bmatrix}.
\end{align}
The stability of this state requires:
\begin{align}
\mathcal{B} \le \mathcal{B}_{c} = \mathcal{K}.
\end{align}

\item \textbf{Collinear state}. In this state, the two N\'{e}el vectors are collinear. It corresponds to the solution:
\begin{align}
\phi_+ = \frac{\pi}{2};
\quad
\phi_- = \pi.
\end{align}
The Hessian matrix reads:
\begin{align}
M = \begin{bmatrix}
\mathcal{B}-\mathcal{K} & 0\\
0 & \mathcal{B}+\mathcal{D}
\end{bmatrix}.
\end{align}
The stability requires:
\begin{align}
\mathcal{B}\ge \mathcal{B}_c = \mathcal{K}.
\end{align}

\item Finally, there is an unstable solution:
\begin{align}
\sin\phi_+ = -\frac{\mathcal{KD}}{\mathcal{B}^2};
\quad
\cos\phi_- = \frac{\mathcal{D}}{\mathcal{B}}.
\end{align}
The Hessian matrix is given by:
\begin{align}
M = \begin{bmatrix}
\mathcal{K}(1-\frac{\mathcal{D}^2}{\mathcal{B}^2}) & \ast \\
\ast & 0
\end{bmatrix}.
\end{align}
The Hessian matrix is not positive semi-definite for all values of $\mathcal{K,D,B}$. Therefore, this solution is unstable over the entire parameter space.
\end{itemize}

To conclude, at low field ($\mathcal{B}<\mathcal{B}_c$), the system is in a non-collinear state which is adiabatically connected to the zero-field state; at high field ($\mathcal{B}<\mathcal{B}_c$), the system enters the collinear state. 

We now compute the magnetization curve. The magnetization parallel ($M^\parallel$) and orthogonal ($M^\perp$) to the magnetic field is given by:
\begin{align}
\frac{M^\parallel}{N} = \frac{(g\mu_B)^2B}{16J_1}(1-\sin\phi_+\cos\phi_-);
\quad
\frac{M^\perp}{N} = -\frac{(g\mu_B)^2B}{16J_1}\cos\phi_+\cos\phi_-.
\end{align}
Using the previously obtained solutions for $\phi_\pm$, we find $M^\perp = 0$ over the entire field range, whereas $M^\parallel$ is given by:
\begin{align}
M^\parallel = \frac{(g\mu_B)^2B}{16J_1}\times\left\{ \begin{array}{cc}
1+ \frac{\mathcal{B}}{\mathcal{K}}& (\mathcal{B}\le\mathcal{B}_c) \\
2 & (\mathcal{B}\ge\mathcal{B}_c) 
\end{array}\right. .
\end{align}

\section{Spin wave gaps of the dipole-biquadratic model}

In this section, we compute analytically the spin wave gaps of the dipole-biquadratic model. We shall see that the value of the gaps is directly related to the strength of the effective dipole and biquadratic interactions. In our calculation, we make the simplifying assumptions that the nearest neighbor Hesienberg exchange interaction $J_1$ is significantly larger than all the other energy scales present in this system.

The Hamiltonian reads:
\begin{align}
H = \sum_{\langle ij\rangle_1}J_1 \mathbf{S}_i\cdot\mathbf{S}_j + \sum_{\langle ij\rangle_3} D(\mathbf{S}_i\cdot\mathbf{S}_j-3(\mathbf{S}_i\cdot\hat{n}_{ij})(\mathbf{S}_j\cdot\hat{n}_{ij}) ) + K(\mathbf{S}_i\cdot\mathbf{S}_j)^2 + A \sum_i (S^z_i)^2.
\end{align}
The first summation over all the nearest neighbor bond describes the dominant intra-layer coupling. The second summation over all the third neighbor bonds describes the weak coupling between the adjacent layers. The last summation over all lattice sites describes the easy plane spin anisotropy. We have omitted $J_2$ and $J_3$ couplings as their contribution to the spin wave gap is second order.

We partition the lattice into four sublattices, which we dubbed as A$_1$, A$_2$. B$_1$, and B$_2$. The spin orientations are given by (in crystallographic axes):
\begin{align}
\mathbf{S}_{A1} = S(1,0,0);
\quad
\mathbf{S}_{A2} = S(-1,0,0);
\quad
\mathbf{S}_{B1} = S(0,1,0);
\quad
\mathbf{S}_{B2} = S(0,-1,0);
\end{align} 
As the lowest energy spin wave excitations are all rigid rotations within the magnetic sublattices, we may set the spins in the same sublattice to take the same orientation. This reduces the Hamiltonian to the following:
\begin{subequations}
\begin{align}
H &= N J_1 (\mathbf{S}_{A1}\cdot\mathbf{S}_{A2} + \mathbf{S}_{B1}\cdot\mathbf{S}_{B2})  + \sum_{\alpha=A_{1,2}}\sum_{\beta=B_{1,2}} H_{\alpha,\beta} + \frac{NA}{4}\sum_{\alpha=A_{1,2},B_{1,2}} (S^z_{\alpha})^2,
\end{align}
where
\begin{align}
H_{\alpha,\beta} &= NK(\mathbf{S}_{\alpha}\cdot\mathbf{S}_{\beta})^2 + ND \left [ \mathbf{S}_{\alpha}\cdot\mathbf{S}_{\beta} - \frac{3c^2}{2a^2+c^2}S^c_{\alpha}S^c_{\beta} - \frac{3a^2}{2a^2+c^2}(S^a_{\alpha}\pm S^b_{\alpha})(S^a_{\beta}\pm S^b_{\beta}) \right].
\end{align}
\end{subequations}
Here, the plus (minus) sign is for the pair A$_1$/B$_1$ and A$_2$/$B_2$ (A$_1$/B$_2$ and A$_2$/B$_1$). $N$ is the number of spins.

In the next step, we choose the local spin frames such that the local $S^z$ axis coincides with the spin direction in the magnetic ground state. We choose the crystallographic $c$ axis as the local $S^x$ axis. We further expand:
\begin{align}
S^{x}_i \approx \sqrt{S} u_i;
\quad
S^{y}_i \approx \sqrt{S} v_i;
\quad
S^{z}_i \approx S-\frac{u^2_i+v^2_i}{2}.
\end{align}
Substituting the above into the Hamiltonian, and expanding to the quadratic order, we obtain:
\begin{align}
H \approx \frac{N}{2} u M^{uu} u + \frac{N}{2} v M^{vv} v,
\end{align}
where the $u$ and $v$ are column vectors:
\begin{align}
u = \begin{bmatrix}
u_{A1} \\
u_{A2} \\
u_{B1} \\
u_{B2}
\end{bmatrix},\quad
v = \begin{bmatrix}
v_{A1} \\
v_{A2} \\
v_{B1} \\
v_{B2}
\end{bmatrix}.
\end{align}
The Hessian matrices read:
\begin{subequations}
\begin{align}
M^{uu} &= S\begin{bmatrix}
J_1+\frac{A}{2}+2\alpha D  & J_1 & -2\beta D & -2\beta D \\
J_1 & J_1+\frac{A}{2}+2\alpha D  & -2\beta D & -2\beta D \\
-2\beta D & -2\beta D & J_1+\frac{A}{2}+2\alpha D  & J_1 \\
-2\beta D & -2\beta D & J_1 & J_1+\frac{A}{2}+2\alpha D 
\end{bmatrix};
\\
M^{vv} &= S\begin{bmatrix}
J_1+4KS^2+2 \alpha D & -J_1 & -2KS^2+\alpha D & -2KS^2+\alpha D \\
-J_1 & J_1+4KS^2+2\alpha D & -2KS^2+\alpha D & -2KS^2+\alpha D \\
-2KS^2+\alpha D & -2KS^2+\alpha D & J_1+4KS^2+2\alpha D & -J_1 \\
-2KS^2+\alpha D & -2KS^2+\alpha D & -J_1 & J_1S+4KS^2+2\alpha D
\end{bmatrix}.
\end{align}
\end{subequations}
Here, we have used short-hand notation:
\begin{align}
\alpha = \frac{3a^2}{2a^2+c^2};\quad \beta =  1-\alpha = \frac{c^2-a^2}{2a^2+c^2}.
\end{align}
The Lagrangian of the model reads:
\begin{align}
L = \frac{N}{4} u^T \dot{v} - H.
\end{align}
The equations of motion are thus:
\begin{align}
\dot{v} = -4M^{uu}u;\quad \dot{u} = 4M^{vv}v.
\end{align}
The spin wave frequencies $\omega_i$ are obtained by diagonalizing the dynamical matrix:
\begin{align}
16\mathrm{Spec}(M^{uu}M^{vv}) = \{\omega^2_i\}.
\end{align}

We are now ready to compute the spin wave frequencies. We note the following unitary transformation
\begin{align}
U = \frac{1}{2}\begin{bmatrix}
1 & 1 & 1 & 1\\
1 & -1 & 1 & -1\\
1 & 1 & -1 & -1\\
1 & -1 & -1 & 1
\end{bmatrix},
\end{align}
\emph{simultaneously} diagonalize the matrices $M^{uu}$ and $M^{vv}$:
\begin{align}
U^\dagger M^{uu} U &= S\begin{bmatrix}
2J_1+\frac{A}{2}+2\alpha D-4\beta D \\
& \frac{A}{2}+2\alpha D \\
& & 2J_1+\frac{A}{2}+2\alpha D + 4\beta D \\
& & & \frac{A}{2}+2\alpha D
\end{bmatrix};
\nonumber\\
U^\dagger M^{vv} U &= S\begin{bmatrix}
4\alpha D  \\
& 2J_1+4KS^2+2\alpha D \\
& & 8KS^2 \\
& & & 2J_1+4KS^2+2\alpha D
\end{bmatrix}.
\end{align}
Thus, the spin wave frequencies are:
\begin{align}
\omega_1 &= 4S\sqrt{(2J_1+\frac{A}{2}+2\alpha D - 4\beta D)4\alpha D} \approx 8S\sqrt{2J_1\alpha D};
\\
\omega_{2,4} &= 4S\sqrt{(2J_1+4KS^2+2\alpha D ) (\frac{A}{2}+2\alpha D)} \approx 4S\sqrt{J_1 (A+\alpha D)};
\\
\omega_{3} &= 4S\sqrt{(2J_1+\frac{A}{2}+2\alpha D + 4\beta D ) 8KS^2} \approx 16S^2\sqrt{J_1 K};
\end{align}

\section{Magnetic dipole interaction}

In this section, we compute the magnetic dipole interaction energy between the two sublattices in LaSrCrO$_4$. Similar to the previous sections, we partition the system into two sublattices, dubbed A and B. The A sublattice consists of the sites with crystallographic label $(i,j,k)$, whereas the B sublattice consists of the sites with label $(i+\frac{1}{2},j+\frac{1}{2},k+\frac{1}{2})$. 

Each of the two sublattices hosts a three-dimensional N\'{e}el order with the N\'{e}el vector lying in the crystallographic $ab$ plane. We parametrize the N\'{e}el vectors by the azimuthal angles $\phi_a$ and $\phi_b$. Here, $\phi_a$ ($\phi_b$) is defined as the azimuthal angle of the spin on site $(0,0,0)$ ($(1/2,1/2,1/2)$).  

We may write the magnetic dipole energy as:
\begin{align}
E_{dip} = E_{AA}+E_{AB}+E_{BB}.
\end{align}
where $E_{00}$ and $E_{11}$ are the intra-sublattice energy and $E_{01}$ the inter-sub-lattice energy. The intra-sublattice energy $E_{00}$ is given by:
\begin{align}
E_{00} &= \frac{N}{4}\frac{\mu_0 m^2}{4\pi a^3}\sideset{}{'}\sum_{ijk} \frac{(-1)^{i+j}}{[i^2+j^2+\eta^2 k^2]^{3/2}} - 3\frac{(-1)^{i+j}(i\cos\phi_a+j\sin\phi_a)^2}{[i^2+j^2+\eta^2 k^2]^{5/2}}
\nonumber\\
& =  \frac{N}{4}\frac{\mu_0 m^2}{4\pi a^3}\sideset{}{'}\sum_{ijk} \frac{(-1)^{i+j}}{[i^2+j^2+\eta^2 k^2]^{3/2}} - \frac{3}{2}\frac{(-1)^{i+j}(i^2+j^2)}{[i^2+j^2+\eta^2 k^2]^{5/2}}.
\end{align}
In the first line, we have used the translation symmetry; in the second line, we have used the symmetry properties to simplify the sum: $\sum ij = 0$, $\sum i^2 = \sum j^2$. The factor of $N/2$ comes from the number of sites in sublattice A; another factor of $1/2$ comes from double counting. $\eta \equiv c/a$ characterizes the lattice geometry. We thus may write $E_{AA}$ as:
\begin{align}
\frac{E_{AA}}{N} = \frac{A}{2}\frac{\mu_0 m^2}{4\pi a^3},
\end{align}
where the numeric constant:
\begin{align}
A = \frac{1}{2}\sideset{}{'}\sum_{ijk} (-1)^{i+j} \frac{-\frac{1}{2}(i^2+j^2)+\eta^2 k^2}{[i^2+j^2+\eta^2 k^2]^{5/2}}.
\end{align}

Inversion symmetry immediate implies $E_{BB} = E_{AA}$. The remaining term is $E_{AB}$:
\begin{align}
E_{AB} &= \frac{N}{2}\frac{\mu_0 m^2}{4\pi a^3} \sum_{ijk} \frac{(-1)^{i+j}\cos(\phi_a-\phi_b)}{[(i-\frac{1}{2})^2+(j-\frac{1}{2})^2+\eta^2 (k-\frac{1}{2})^2]^{3/2}} 
\nonumber\\
&- 3\frac{(-1)^{i+j} [(i-\frac{1}{2})\cos\phi_a+(j-\frac{1}{2})\sin\phi_a] [(i-\frac{1}{2})\cos\phi_b+(j-\frac{1}{2})\sin\phi_b] }{[(i-\frac{1}{2})^2+(j-\frac{1}{2})^2+\eta^2 (k-\frac{1}{2})^2]^{5/2}}
\nonumber\\
& = \frac{N}{2}\frac{\mu_0 m^2}{4\pi a^3} \times -3\sum_{ijk} \frac{(-1)^{i+j} (i-\frac{1}{2})(j-\frac{1}{2})\sin(\phi_a+\phi_b) }{[(i-\frac{1}{2})^2+(j-\frac{1}{2})^2+\eta^2 (k-\frac{1}{2})^2]^{5/2}}.
\end{align}
In the first line, we have used the translation symmetry. The factor of $N/2$ comes from the number of sites in sublattice 0. There is no double counting factor. In the second line, we have used the symmetry properties: $\sum (-1)^{i+j} = 0$, $\sum (-1)^{i+j}(i-1/2)^2 = 0$, etc. We thus write:
\begin{align}
\frac{E_{AB}}{N} = A'\frac{\mu_0 m^2}{4\pi a^3}\sin(\phi_a+\phi_b),
\end{align}
where the numeric constant
\begin{align}
A' = -\frac{3}{2}\sum_{ijk} (-1)^{i+j} \frac{(i-\frac{1}{2})(j-\frac{1}{2})}{[(i-\frac{1}{2})^2+(j-\frac{1}{2})^2+\eta^2 (k-\frac{1}{2})^2]^{5/2}}.
\end{align}

Using the lattice constants $a = 3.872$\,{\AA} and $c = 12.516$\,{\AA}, a direct numerical summation of the series reveals $A' = -4.2484\times10^{-2}$. Taking the experimentally measured static moment $m = 2.25\mu_B$, we obtain the coupling:
\begin{align}
\frac{E_{AB}}{N} = - 1.9891\times10^{-4} \sin(\phi_a+\phi_b)\,(\mathrm{meV}).
\end{align} 
Meanwhile, using the dipole-biquadratic model, we find the coupling between the two sublattices read:
\begin{align}
\frac{E_{AB}}{N} = -\frac{12a^2}{2a^2+c^2}DS^2\sin(\phi_a+\phi_b) = -0.964 DS^2\sin(\phi_a+\phi_b).
\end{align}
Comparing the two, we obtain:
\begin{align}
DS^2 =2.063\times 10^{-4}\,\mathrm{meV}.
\end{align}

\section{Spin flop transitions of the dipole-anisotropy model}

In this section, we discuss the spin flop transitions of the dipole-anisotropy model. When the field is along the $[100]$ direction, the model exhibits a first order spin flop transition, at which the magnetization curve shows a sudden jump. This is inconsistent with the experimental results.

\begin{figure}
\centering
\includegraphics[width=0.8\textwidth]{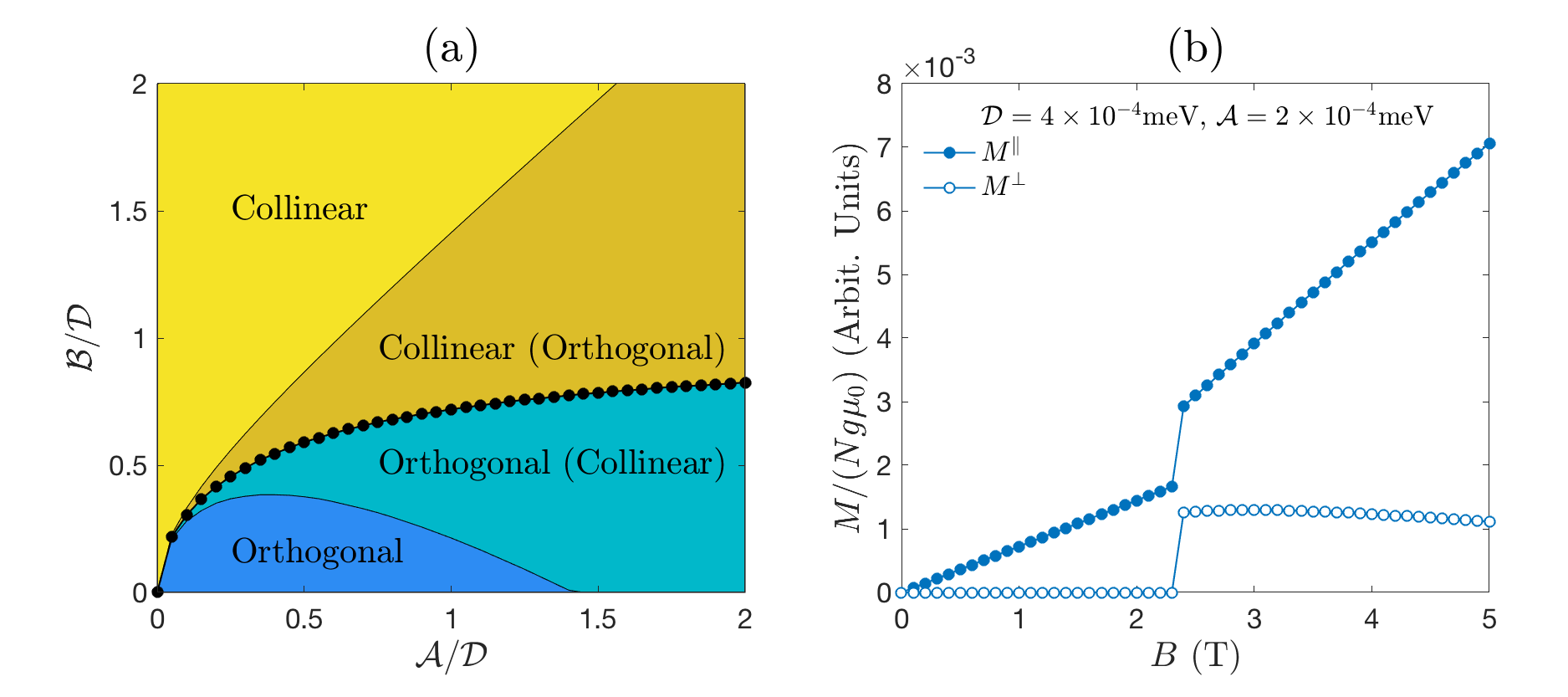}
\caption{(a) Spin flop transitions of the D-A model in $[100]$ field. (b) Magnetization curve for typical model parameters.}
\end{figure}

The analysis of the spin flop transitions of the dipole-anisotropy model is similar to that of the dipole-biquadratic model. We consider the case with the field $\parallel [100]$. Again, we divide the system into A and B sublattices, each hosts a three-dimensional N\'{e}el order parametrized by the azimuthal angle $\phi_a$ ($\phi_b$). The total energy reads:
\begin{align}
\frac{E_\mathrm{tot}}{N} = -\frac{(g\mu_0B)^2}{32J_1}(\sin^2\phi_a+\sin^2\phi_b)-\frac{12a^2}{2a^2+c^2}D\sin(\phi_a+\phi_b)+\frac{A' S^4}{8}(\sin^2 2\phi_a+\sin^2 2\phi_b).
\end{align}
Here, the first term is the energy due to the magnetic field; the second the dipolar interaction; and the third the magnetocrystalline spin anisotropy. We define $\phi_+ = \phi_a+\phi_b$ and $\phi_- = \phi_a-\phi_b$, and rewrite the energy as:
\begin{align}
\frac{E_\mathrm{tot}}{N} = \mathcal{B}\cos\phi_+\cos\phi_- -\mathcal{D}\sin\phi_+ -\frac{\mathcal{A}}{4}\cos2\phi_+\cos2\phi_-,
\end{align}
where $\mathcal{B,D}$ have been defined before, and
\begin{align}
\mathcal{A} = \frac{A' S^4}{2},
\end{align}
is the energy scale of the spin anisotropy energy. The stationary condition reads:
\begin{align}
\sin\phi_- (-\mathcal{B}\cos\phi_+ + \mathcal{A}\cos2\phi_+ \cos\phi_-) = 0;
\quad
\sin\phi_+ (-\mathcal{B}\cos\phi_- +\mathcal{A}\cos2\phi_-\cos\phi_+)-\mathcal{D}\cos\phi_+ = 0.
\end{align}
The Hessian matrix reads:
\begin{align}
M = \begin{bmatrix}
-\mathcal{B}\cos\phi_+\cos\phi_- + \mathcal{A}\cos2\phi_+\cos2\phi_- & \mathcal{B}\sin\phi_+\sin\phi_- -\mathcal{A}\sin2\phi_+\sin2\phi_- \\
\mathcal{B}\sin\phi_+\sin\phi_- -\mathcal{A}\sin2\phi_+\sin2\phi_- & -\mathcal{B}\cos\phi_+\cos\phi_- +\mathcal{D}\sin\phi_+ +  \mathcal{A}\cos2\phi_+\cos2\phi_- 
\end{bmatrix}.
\end{align}

We find two locally stable solutions:
\begin{itemize}
\item \textbf{Orthogonal state}. This state is identical to that of the zero-field state. It corresponds to the following solution:
\begin{align}
\phi_+ = \frac{\pi}{2};\quad
\phi_- = \pm\frac{\pi}{2}
\end{align}
The Hessian matrix reads:
\begin{align}
M = \begin{bmatrix}
\mathcal{A} & \pm\mathcal{B} \\
\pm\mathcal{B} & \mathcal{A}+\mathcal{D}
\end{bmatrix}.
\end{align}
The stability requires:
\begin{align}
\mathcal{B}^2 \le \mathcal{B}^2_{s1} = \mathcal{A}(\mathcal{A}+\mathcal{D}). 
\end{align}
The energy of this state is:
\begin{align}
\frac{E_\mathrm{tot}}{N} = -\mathcal{D}-\frac{\mathcal{A}}{4}.
\end{align}

\item \textbf{Collinear state}. In this state, the N\'{e}el vectors are collinear. The angles are determined by the following transcendental equation:
\begin{align}
\phi_- = 0;\quad
\mathcal{A}\sin\phi_+\cos\phi_+ = \mathcal{D}\cos\phi_+ + \mathcal{B}\sin\phi_+.
\end{align}
We define $\mathcal{B} = \sqrt{\mathcal{B}^2+\mathcal{D}^2} \cos\theta$ and $\mathcal{D} = \sqrt{\mathcal{B}^2+\mathcal{D}^2} \sin\theta$. The equation now reads:
\begin{align}
\sin(\phi_+ + \theta) = \frac{\mathcal{A}}{\sqrt{\mathcal{B}^2+\mathcal{D}^2}}\sin\phi_+ \cos\phi_+  .
\end{align}
Numerically, it is much more convenient to seek the minimum of the energy function:
\begin{align}
\frac{E_\mathrm{tot}}{N} = \sqrt{\mathcal{B}^2+\mathcal{D}^2} \cos(\phi_+ +\theta)-\frac{\mathcal{A}}{4}\cos2\phi_+.
\end{align}
The Hessian matrix reads:
\begin{align}
M = \begin{bmatrix}
-\mathcal{B}\cos\phi_+ + \mathcal{A}\cos2\phi_+ & 0\\
0 & -\mathcal{B}\cos\phi_+ + \mathcal{A}\cos2\phi_+ + \mathcal{D}\sin\phi_+
\end{bmatrix}.
\end{align}
We expect that the collinear state is stable when $\mathcal{B}\ge \mathcal{B}_{s2}$, where $\mathcal{B}_{s2}$ must be determined numerically. 
\end{itemize}

We have numerically determined the value of $\mathcal{B}_{s1}$ and $\mathcal{B}_{s2}$ and mark the region of local stability for both orthogonal and collinear states. We find that $\mathcal{B}_{s1}\le \mathcal{B}_{s2}$. This implies that there is a region where both local minima are stable --- this indicates a first order transition. We therefore must determine the transition by comparing the energy at these local minima. 

The magnetization:
\begin{align}
\frac{M^\parallel}{N} &= \frac{(g\mu_0)^2B}{16J_1}\times\left\{
\begin{array}{cc}
1 & (\mathcal{B} \le \mathcal{B}_{c}) \\
1-\cos\phi_+ & (\mathcal{B}_{c} \le \mathcal{B})
\end{array}\right. ;
\\
\frac{M^\perp}{N} &= \frac{(g\mu_0)^2B}{16J_1}\times\left\{
\begin{array}{cc}
0 & (\mathcal{B}\le \mathcal{B}_{c}) \\
\pm\sin\phi_+ & (\mathcal{B}_{c} \le \mathcal{B})
\end{array}\right. 
\end{align}
The angle $\phi_+$ has to be determined by solving the equation.

\end{document}